\documentclass[twocolumn,trackchanges]{aastex701}

\newcommand{\msini}{m\,\sin i_{\rm orb}}

\newcommand{\snrast}{\text{SNR}_\text{ast}}

\usepackage{amsmath}
\usepackage{textcomp}

\begin{document}

\title{CLS-Gaia. I. Expected Results from a Joint Analysis of Time Series Radial Velocities from the California Legacy Survey and Astrometry from Gaia-DR4}

\author[0000-0002-9305-5101]{Luke B. Handley}
\altaffiliation{NSF Graduate Research Fellow}
\affiliation{Department of Astronomy, California Institute of Technology, Pasadena, CA 91125, USA}
\email[show]{lhandley@caltech.edu}  

\author[0000-0001-8638-0320]{Andrew W. Howard}
\affiliation{Department of Astronomy, California Institute of Technology, Pasadena, CA 91125, USA}
\email{ahoward@caltech.edu}

\author[0000-0002-1838-4757]{Aniket Sanghi}
\altaffiliation{NSF Graduate Research Fellow}
\affiliation{Department of Astronomy, California Institute of Technology, Pasadena, CA 91125, USA}
\email{asanghi@caltech.edu}

\author[0000-0002-8895-4735]{Dimitri Mawet}
\affiliation{Department of Astronomy, California Institute of Technology, Pasadena, CA 91125, USA}
\affiliation{Jet Propulsion Laboratory, California Institute of Technology, Pasadena, CA 91109, USA}
\email{dmawet@astro.caltech.edu}

\begin{abstract}
Most of the known giant planets orbiting within $\sim$30 au of their host stars were discovered by Doppler surveys. The California Legacy Survey (CLS) is a benchmark planet survey, given its large sample size (719 nearby bright stars with 178 detected planets), long observational baseline ($\sim$20 years), and relatively high precision ($\sim$2\,m\,s$^{-1}$).
Radial velocities (RVs) measure many orbital quantities precisely but are insensitive to orbital inclination ($i_\mathrm{orb}$) and place only lower limits on planet masses ($\msini$).
These shortcomings can be overcome by combining the CLS RVs with time-series astrometry from the fourth data release (DR4) of the Gaia mission.
In this paper, we describe a methodology for simultaneously fitting astrometry and RVs and forecast the measurement of full 3D orbits for $\sim$80 known giant planets from CLS, double the expected yield of an astrometry-only search for those stars, with masses and orbital inclinations measured to $\sim$10\% precision.  The sample is especially valuable because the planets are well-studied, independently verified, and their masses extend below a Jupiter mass, whereas most Gaia discoveries will be super-Jupiters and brown dwarfs.
Among the most interesting planetary systems in CLS are those with two or more giant planets.  We forecast that 16 of these multi-giant systems will have measurable planet-planet mutual inclinations, while an astrometry-only analysis would yield no such measurements for the CLS stars. The simulated data are sufficient to distinguish between plausible mutual inclination distributions, including coplanar like Jupiter and Saturn, a broad range of mutual inclinations, and an aligned–isotropic mixture.

\end{abstract}


\section{Introduction} 

Among the demographic surveys for extrasolar planets, the California Legacy Survey (CLS) stands out for its sensitivity to giant planets and brown dwarfs with semimajor axes from 0.01 au to $\sim$30 au around a large sample of ``sun-like stars'' (late F dwarfs to mid-M dwarfs).  In total, 719 nearby, bright stars were surveyed and 178 planets were detected \citep{Rosenthal2021}, with many discoveries made by other groups and confirmed in the CLS measurements. The RV measurements were made using HIRES at the W.\,M.\,Keck Observatory \citep{Vogt1994} during 1996--2020, the Hamilton spectrograph at Lick Observatory \citep{Fischer2014} during 1987--2012, and the Levy spectrograph \citep{Vogt2014} during 2012--2020.  The target stars were selected for CLS without knowledge of whether they host planets.  While the RVs do not have the uniform observational cadence of a space mission, minimum standards were imposed on the number of RVs and their observational baseline so that demographic analyses of the CLS sample are sensitive to the intrinsic distribution of planets and their characteristics. 

The first CLS paper detailed target selection and publicized the RV dataset and the \texttt{RVSearch} methodology for planet search, and reported 14 newly confirmed planets \citep{Rosenthal2021}. Subsequent studies characterized giant-planet occurrence rates \citep{Fulton2021}, the relationship between outer giant planets and inner smaller planets \citep{Rosenthal2022}, and the distinct metallicity and eccentricity distributions of systems hosting multiple giant planets \citep{Rosenthal2024}. Recent CLS studies have also explored the planet--brown dwarf boundary through metallicities, eccentricities, and occurrence rates \citep{Giacalone2026,Gilbert2026,VanZandt2026}. Together, these and other studies have established a detailed demographic picture of nearby giant planets, but one based primarily on their projected orbital properties.

RVs measure the line-of-sight component of stellar reflex motion, constraining the minimum planetary mass $m\sin i_\mathrm{orb}$ but not the orbital inclination $i_\mathrm{orb}$ itself. Consequently, RVs alone cannot measure true planetary masses or the three-dimensional architectures of planetary systems. Astrometry provides a complementary observable by measuring stellar positions in the plane of the sky. Joint RV/astrometric modeling can therefore constrain three-dimensional orbits and recover true dynamical masses.  In systems of multiple planets, measurements of the orbital inclinations and longitudes of ascending nodes can be used to measure mutual inclinations between planets. The Gaia mission \citep{Lindegren2012,Prusti2016,Brown2016,Brown2018,Vallenari2023} provides precise astrometry for billions of stars up to brightnesses of $G\approx20$, including all CLS targets.

This paper introduces the CLS-Gaia project, which combines the CLS RV measurements with the forthcoming Gaia astrometric time series. The $\sim$5.5-year time series of astrometric measurements in Gaia DR4 (2014--2019) has the highest expected sensitivity for planets at $\sim$0.5--3 au. The techniques in CLS-Gaia involve modeling of the time series RVs and astrometry with Keplerian models (and N-body models for one planetary system) to measure full, three-dimensional orbits and true masses of planets and brown dwarfs.  Orbital inclination on the sky is an intrinsically uninteresting angle because it depends on our arbitrary viewing geometry.  However, in systems of multiple planets and brown dwarfs, the mutual inclination between two three-dimensional orbits carry physically meaningful information.  CLS-Gaia will add these two elements---mutual inclinations and true masses---to more fully characterize the planetary systems. 

Ultimately, CLS-Gaia will enable a follow-up demographic analysis of the CLS population, revisiting planet occurrence as a function of mass and orbital separation with astrometrically-constrained true masses \citep{Fulton2021,VanZandt2026}. This will include a search for additional planets in both the RV and Gaia data, although that is outside the scope of this paper.

The scientific focus of this paper is planetary systems of two or more giant planets. Outside the specific archetype of compact multi-transiting planetary systems, planet-planet inclinations are scarce in the literature \citep[see, e.g.,][]{Xiao2026a}, limiting constraints on the underlying population and its selection effects\footnote{For example, the Hubble Space Telescope measurements of $\nu$~And and GJ~876 yielded orbital configurations that have proven difficult to reconcile with subsequent dynamical and astrometric constraints \citep{McArthur2010,Benedict2002,Benedict2022}.}. A homogeneous sample is the best path towards distinguishing a tightly coplanar population (like those observed in Kepler; \citealt{Xie2016,Fabrycky2014}), dynamically excited systems shaped by scattering and relaxation (perhaps similar to the eccentricities of giant planets; \citealt{Juric2008,Ford2008}), and highly inclined systems that may facilitate high-eccentricity migration \citep{Nagasawa2008,Naoz2011}. Thus, measuring these angles may mean constraining the formation and dynamical evolution of long-period giant-planet systems.

The remainder of this paper is organized as follows. Section~\ref{sec:intuition} provides an introduction to the relevant astrometric observables and their implications for CLS. Section~\ref{sec:Methods} describes our methodology for fitting RVs and astrometry simultaneously.  Section~\ref{sec:joint_fitting} has a benchmark comparison of our methods with those of \citet{Rosenthal2021}, evaluates the results using simulated Gaia DR4 astrometry, and forecasts the resulting orbital constraints. Section~\ref{sec:results} identifies the CLS systems for which a three-dimensional orbital characterization is likely, and Section~\ref{sec:discussion} explores the scientific applications of the proposed measurements. Section~\ref{sec:conclusion} offers conclusions.

\section{Astrometric Detection} \label{sec:intuition}

Realizing the potential of CLS-Gaia depends on the astrometric data products available from Gaia. Gaia DR3, which did not provide astrometric measurements at individual epochs, has already demonstrated the promise of astrometry for exoplanet discovery and characterization. This includes a catalog of exoplanet candidates \citep{Holl2023} and several systems amenable to direct comparison with RV observations. Previous studies have likewise combined Gaia astrometry with RVs to characterize individual giant planets \citep{Li2021,Winn2022}, although in many cases the companion's orbit is inferred indirectly from the available data (e.g., \citealt{Kiefer2019}), or from changes in the bulk astrometric motion between the Gaia and Hipparcos missions \citep{Perryman1997,Brandt2018,Brandt2021}. The principal advance of DR4 will be the availability of the underlying astrometric time series, rather than solely fitted single-star or binary solutions. In isolation, these time series are expected to yield thousands of new exoplanet candidates \citep{Perryman2014,Lammers2026}. The bulk statistical yield of DR4 is the subject of other work, and we do not attempt to reproduce it here because the CLS-Gaia project relies (mostly) on previously detected planetary systems using RVs.

Joint modeling of RV and astrometric time series is described in, e.g., \citet{Sahlmann2011}, for Hipparcos epoch astrometry. The astrometric precision of Hipparcos restricted its applicability primarily to M-dwarf and brown-dwarf companions ($\sim100\,M_J$). Gaia substantially extends this regime. A useful metric for this improvement is the angular scale of the orbit on the sky, the astrometric signature $\alpha_\textrm{ast}$, defined by \citet{Perryman2014} as
\begin{equation} \label{eq:signature}
    \alpha_\textrm{ast} = \left(\frac{m}{M_\star}\right)\left(\frac{a}{1\,\text{au}}\right) \left(\frac{d}{1\,\text{pc}}\right)^{-1} \text{arcsec},
\end{equation}
where $m$ is the companion mass, $M_\star$ is the host mass, $a$ is the orbital semimajor axis, and $d$ is the distance to the system. The signature $\alpha_\textrm{ast}$ scales linearly with companion mass, and Gaia astrometry is  $\sim$100$\times$ more precise per-visit than Hipparcos (with some dependence on source magnitude; \citealt{Perryman1997,Perryman2014,ElBadry2024}). Extrapolating from $\alpha_\textrm{ast}$ alone, one would expect joint RV/astrometric modeling to reach into the giant-planet regime, at roughly $1\,M_J$ for nearby stars, for the first time. 

\begin{figure}
    \centering
    \includegraphics[width=1.0\linewidth]{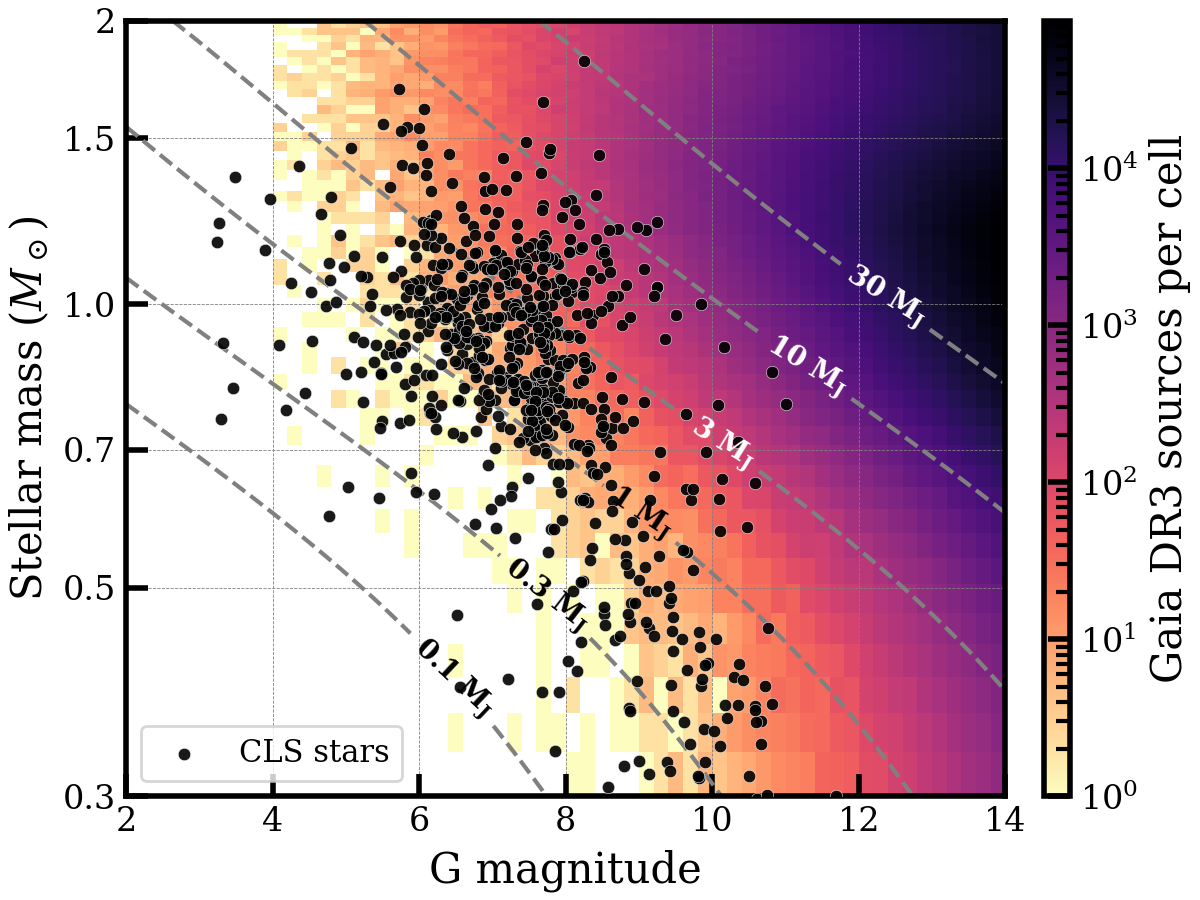}
    \caption{The CLS stellar sample (black points) in the magnitude–stellar-mass plane, with Gaia DR3 stars possessing reliable parallax measurements shown as shaded cells for comparison. Contours of the maximum detectable companion mass at 1~au are shown. For consistency in this figure, the stellar masses of both samples are estimated from the Gaia DR3 G-band magnitude and distance, using a fifth-order polynomial fit to the piecewise stellar mass–luminosity relation of \citet{Eker2018}.}
    \label{fig:cls_and_gaia}
\end{figure}

CLS stars are among the nearest in the Galaxy. Figure \ref{fig:cls_and_gaia} shows the CLS sample in the plane of G-band magnitude and stellar mass, which overlap with the most favorable regions for astrometric detection. Because of their brightness and proximity, CLS stars benefit from favorable Gaia astrometric precision, with a single-epoch astrometric uncertainty of $\sigma_\eta\approx54~\mu$as for stars with $\mathrm{G}\lesssim13$ (with some degradation for the brightest stars, especially when $G<4$; \citealt{ElBadry2024}). \citet{Lammers2026} found that astrometric detections correlate with a minimum per-epoch signal-to-noise ratio (SNR) of $\alpha_\textrm{ast}/\sigma_\eta\approx1.5$. The figure shows contours corresponding to this threshold for companions of various masses on 1~au orbits. In contrast to the CLS sample, the vast majority of stars observed by Gaia, including those brighter than $\mathrm{G}=14$ shown in Figure \ref{fig:cls_and_gaia}, are not sufficiently sensitive to detect companions with masses $\lesssim10M_\mathrm{J}$ in DR4.

\begin{figure}
    \centering
    \includegraphics[width=1.0\linewidth]{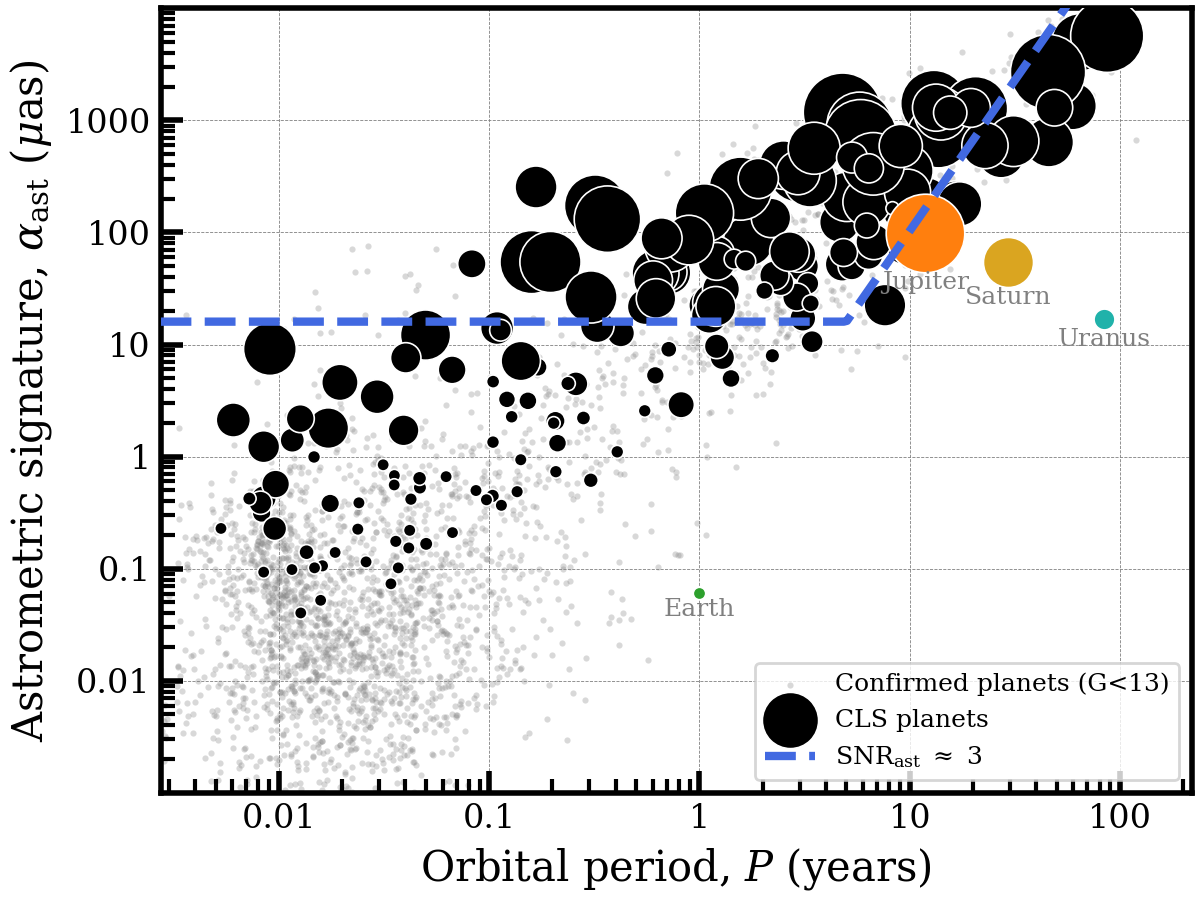}
    \caption{Astrometric signatures of the 178 CLS planets, with point sizes scaled by minimum mass. The signatures of Jupiter, Saturn, Uranus, and Earth, are shown for an observer situated 50 pc from the Solar System. Known exoplanets orbiting stars with $G<13$ (drawn from the NASA Exoplanet Archive on 16 November 2026 \citep{Christiansen2025,PSCompPars}) are shown as small gray dots.  The blue dashed line denotes the sensitivity contour for  $\snrast\approx 3$ (Equation \ref{eq:snr}).}
    \label{fig:cls_planets}
\end{figure}

Of course, the subset of systems likely to benefit from Gaia astrometric constraints also depends on the properties of the planets therein. Figure \ref{fig:cls_planets} shows the amplitudes of known CLS planets as functions of orbital period (compare with Figure 1 in \citealt{Perryman2014}). Generally, the planets known to CLS complement the regions of parameter space where Gaia is most sensitive --- giant planets orbiting at au scales. 

The widest planetary systems, regardless of their astrometric amplitudes, are still unlikely to be detected with DR4 astrometry. Thus far, we have not taken into account other factors that impact the quality of the astrometric fit, including the number of observations (depending on galactic coordinates) and the extent to which distinct phases of the orbit are sampled over the observing baseline. In the remainder of this paper, we will stray away from use of the per-epoch SNR, and instead make use of the total astrometric contribution,
\begin{equation} \label{eq:snr}
    \snrast \equiv
    \frac{\alpha_\textrm{ast}}{\sigma_\eta}\sqrt{N_j} \times \mathcal{F}(P),
\end{equation}
where $N_j$ is the number of astrometric epochs and $P$ is the companion's orbital period. The factor $\mathcal{F}(P)$ accounts for sensitivity losses when the orbital period extends beyond the Gaia DR4 baseline:
\begin{equation} \label{eq:period_penalty}
    \mathcal{F}(P) \approx
    \begin{cases}
        \left(\dfrac{P}{5.5\,\text{yr}}\right)^{-2}, 
        & P > 5.5\,\text{yr} \\[2mm]
        1, & \text{otherwise,}
    \end{cases}
\end{equation}
where the slope of $-2$ comes from an empirical calibration (see Section \ref{sec:results}). As we will show, astrometry contributes to a joint RV/astrometry fit when  $\snrast\approx3$. A contour of this value in $\alpha_\textrm{ast}$-$P$ space is also shown in Figure \ref{fig:cls_planets}, emphasizing the limitation. 

\section{Methodology} \label{sec:Methods}

\subsection{Principles}

\label{sec:principles}

The RV method of planet detection is sensitive to only a subset of a planet's orbital degrees of freedom that contribute directly to line-of-sight motion. For each planet in a given system, the RV measurements ${v_i}$ at epochs ${t_i}$ constrain the orbital period $P$, time of pericenter passage $T_p$, orbital eccentricity $e$, argument of pericenter $\omega$, and the RV semi-amplitude $K$. If the host star's mass is well known, the semi-amplitude can be converted to the projected planet mass $\msini$ via
\begin{equation} \label{eq:semiamplitude}
    K = \left(\frac{2 \pi G}{P}\right)^{1/3}\frac{\msini}{(M_\star+m)^{2/3}}\frac{1}{ \sqrt{1-e^2}},
\end{equation}
assuming a Keplerian approximation for the orbital solution (see Section \ref{sec:nbody}).

The bulk properties of the CLS stars are well characterized in a homogeneous spectral analysis using \texttt{SpecMatch} \citep{Petigura2015,Petigura2017} and \texttt{isoclassify} \citep{Huber2017,Berger2020,Berger2023}. The former measures each star's effective temperature $T_{\mathrm{eff}}$, surface gravity $\log g$, and metallicity $\mathrm{[Fe/H]}$ by comparison to a synthetic spectral library. The latter estimates stellar mass $M_\star$ and radius $R_\star$ (among other quantities) by interpolation over isochrone grids. \citet{Rosenthal2021} reported a median CLS stellar mass uncertainty of $\sim$3\% from this pipeline, making the conversion from $K$ to $\msini$ straightforward. The orbital parameters that cannot be recovered from RVs alone are therefore just three: the orbital inclination $i_{\mathrm{orb}}$, the true planet mass $m$, and the longitude of ascending node $\Omega$.

Astrometry measures the plane-of-sky component of the orbital motion. Gaia records stellar positions at specific epochs along the time-varying orientation of its CCD mosaic \citep[see, e.g., Figure 4 of][]{Prusti2016}, with each `along-scan' (AL) measurement projected onto celestial coordinates using its precise epoch $\tau_j$ and orbital solution. Variations in these coordinates trace the star's motion across the sky, including the effects of parallax, proper motion, and periodic perturbations arising from bound companions.

Unlike the CLS RVs, Gaia epoch astrometry can constrain the full set of orbital elements (still subject to some degeneracies) from a single dataset. However, the relatively short DR4 baseline of $5.5$ years limits sensitivity to longer-period companions, and the astrometric sensitivity limits Gaia to characterization of mostly giant planets (and larger). When an orbit is not fully resolved, the observed short astrometric arc is consistent with a wide range of companion orbits and masses, limiting the standalone DR4 exoplanet yield to $\sim3$ au \citep{Lammers2026}. Thus, Gaia alone is insensitive to some of the orbital range where giant planets are the most common \citep{Fulton2021,VanZandt2026}.

A second complication is that Gaia measures the system's photocenter rather than the central star itself \citep{Winn2022}. The observed angular semimajor axis scales as
\begin{equation} \label{eq:epsilon}
a_{\mathrm{obs}} \propto \left| \frac{\epsilon}{1+\epsilon} - \frac{q}{1+q}\right|,
\end{equation}
where $\epsilon$ and $q$ are the secondary-to-primary luminosity and mass ratios, respectively. This degeneracy can produce substantial false-positive rates among Gaia planet candidates \citep{Marcussen2023,Stefansson2025} and will complicate the interpretation of future releases \citep{Lammers2026}.  For CLS planets that have been validated by RV characterization, false positives are not a problem. In addition, the complementary information provided by RVs can constrain the companion mass and thereby  disentangle the astrometric signal.

\subsection{Parameters}

With the importance of a joint model motivated, we now turn to the technical details of our fitting pipeline. The parameters associated with a star of mass $M_\star$ and its astrometric motion are denoted by
\begin{equation}
    \theta^{\star} = \{ M_\star, \alpha,\delta, \mu_\alpha, \mu_\delta , \varpi\}
\end{equation}
where $\alpha$ and $\delta$ are the right ascension and declination, $\mu_\alpha$ and $\mu_\delta$ are their respective proper motions, and $\varpi$ is the system parallax. These five astrometric terms mirror Gaia's standard single-star solution (the ``five-parameter fit''), allowing our model to absorb any motion not attributed to bound companions. 

Additional parameters describe the mass, brightness, and orbit of a companion: 
\begin{equation}
    \theta^{\text{orb}} = \{ K,\log P,\phi,\cos i_{\text{orb}},\Omega,\sqrt{e} \cos \omega,\sqrt{e} \sin \omega,\log\epsilon \}.
\end{equation}
The quantities $e$ and $\omega$ are parameterized as $\sqrt{e}\cos\omega$, $\sqrt{e}\sin\omega$ so that the pair can be sampled uniformly over a disk, avoiding bias toward high eccentricities. Similarly, $\cos i_{\text{orb}}$ is sampled rather than $i_{\text{orb}}$ itself, so that orbital inclinations are drawn uniformly over a sphere rather than biased toward edge-on configurations. The phase of the orbit at a reference epoch\footnote{In this work, we adopted a reference epoch halfway between the first and last dates covered by DR4, although the choice is arbitrary.} $\phi$ is used in place of $T_p$ to transform an otherwise periodic posterior into the finite interval $[0,1)$. The flux ratio $\epsilon$ enters as in Equation~\ref{eq:epsilon} and allows the astrometric amplitude to depend on the companion's luminosity as well as its mass. During fitting, the RV semi-amplitude $K$ is given a lower bound of zero\footnote{We note that this choice bears risk of biasing $K$ towards higher values. However, among the 76 planets detected by the joint model (Section \ref{sec:results}), only one planet was detected below 7$\sigma$ significance in $K$ --- HD~181234~b was detected at 3.7$\sigma$ significance. Even at this level, the probability density near an amplitude of zero is negligible, making the bias insignificant.} to avoid nonphysical solutions.

The model also includes nuisance parameters that account for underestimated measurement errors (especially those that are instrumental and astrophysical in origin). The jitter term $\sigma_{\text{ast}}$ accounts for unmodeled systematics in the epoch astrometry.  Corresponding RV jitter terms $\sigma_{\text{RV,k}}$ are included the $N_k$ RV instruments (up to four, as HIRES data before and after the 2004 detector upgrade are modeled independently). Jitter terms are sampled in log-space so that they can span large ranges. The RV instruments have independent RV zero points $\gamma_k$.  Slow change in RV over time (e.g., due to a massive companion with a long orbital period) are modeled with linear ($\dot \gamma$) and quadratic ($\ddot \gamma$) terms.

These terms, however, can become problematic in systems hosting long-period planets, where a real orbital signal can be partially absorbed by the polynomial trend and bias the recovered parameters. We found a reliable threshold for including higher-order polynomial terms to be when every planet in the system has an orbital period below $\approx 60\%$ of the observing baseline, and we adopted this convention throughout this work.

Combining the above terms, the instrumental fitting vector is
\begin{equation}
    \theta^{\rm inst}
    =
    \left\{
        \{\gamma_k,\log \sigma_{\textrm{RV},k}\}_{k=1}^{N_k},\dot{\gamma},\ddot{\gamma},
        \log \sigma_{\rm ast}
    \right\},
\end{equation} 
where $\dot \gamma$ and $\ddot \gamma$ may be omitted. The complete fitting vector $\theta^{\text{sys}}$ is then
\begin{equation}
    \theta^{\text{sys}} = \left\{ \theta^\star ,  \theta^{\rm inst}, \theta^{\rm orb}_1 \ldots,\theta^{\rm orb}_{N_c} \right\},
\end{equation}
where $N_c$ is the number of companions to that star.

\subsection{Model}

The log-likelihood of our model is taken to be the sum of the log-likelihoods of the RV and astrometric models,
\begin{equation}
    \ln \mathcal{L}_{\text{tot}} = \sum_{k=1}^{N_k} \ln \mathcal{L}_{\text{RV}}^k \, + \, \ln \mathcal{L}_{\text{ast}},
\end{equation}
where $\mathcal{L}_{\text{RV}}^k$ is the likelihood of $\theta_{\text{sys}}$ given the RV data from instrument $k$ alone, and $\mathcal{L}_{\text{ast}}$ is the likelihood of $\theta_{\text{sys}}$ given the epoch astrometry alone. We define both explicitly below.

For all parameters except $M_\star$, for which we impose a Gaussian prior based on the spectroscopic mass constraint from the CLS catalog, uninformative priors are adopted. The sampling parameters $\sqrt{e}\cos\omega$ and $\sqrt{e}\sin\omega$ are constrained to the unit disk, and $\cos i_{\text{orb}}$ is constrained to the uniform interval $[-1,1]$.

\subsubsection{RV Likelihood} \label{sec:rvmodel}

The RV log-likelihood is taken to be
\begin{align}
    \ln \mathcal{L}_{\text{RV}}^k = -\frac{1}{2}\sum_{i=1}^{N_i^k}
    \bigg[
        &\frac{\left(v_i - f(t_i,\theta)\right)^2}{\sigma_{v,i}^2 + \sigma_{\text{RV},k}^2} \notag \\
        &+ \ln\!\left(2\pi\left(\sigma_{v,i}^2 + \sigma_{\text{RV},k}^2\right)\right)
    \bigg],
\end{align}
where $f(t_i,\theta)$ is the model RV (plus any polynomial trend, where applicable) evaluated at $t_i$ for the system parameters $\theta_{\rm sys}$, $\sigma_{v,i}$ is the reported measurement uncertainty, and $\sigma_{\text{RV},k}$ is the jitter term for instrument $k$, added in quadrature. The sum runs over the $N_i^k$ epochs observed by instrument $k$.

\subsubsection{Astrometric Likelihood}

The astrometric likelihood is constructed analogously, using the AL formalism introduced in Section~\ref{sec:principles}. At each epoch $\tau_j$, Gaia records a single AL abscissa residual $\eta_j$, the star's displacement projected onto the scan-angle direction $\psi_j$ at that epoch. The model predicts this displacement as
\begin{equation}
g(\tau_j,\theta) = \Pi_j\varpi + \bar\alpha^*(\tau_j) \sin\psi_j + \bar\delta(\tau_j)\cos\psi_j,
\end{equation}
where $\Pi_j$ is the parallax factor at epoch $\tau_j$ (the projection of the Earth--Sun vector onto the scan direction $\psi_j$) and $\varpi$ is the star's parallax (from $\theta^\star$). The quantities $\bar\alpha^*(\tau_j)$ and $\bar\delta(\tau_j)$ are the total predicted offsets in right ascension\footnote{In the same convention as the Gaia team, we use $\bar \alpha^\star$ as shorthand for the projection onto the celestial sphere, $\bar \alpha \cos\delta$, rather than the raw coordinate offset $\bar\alpha$.} and declination at epoch $\tau_j$, summing the linear proper motion and the reflex orbital motion of the photocenter induced by each companion. The latter are computed using the Thiele-Innes formalism (see Section \ref{sec:astro_simulation}). The corresponding log-likelihood is then
\begin{align}
    \ln \mathcal{L}_{\text{ast}} = -\frac{1}{2}\sum_{j=1}^{N_j}
    \Bigg[
        &\frac{\left(\eta_j - g(\tau_j,\theta)\right)^2}{\sigma_{\eta,j}^2 + \sigma_{\text{ast}}^2} \notag \\
        &+ \ln\!\left(2\pi\left(\sigma_{\eta,j}^2 + \sigma_{\text{ast}}^2\right)\right)
    \Bigg],
\end{align}
where $\sigma_{\eta,j}$ is the AL measurement uncertainty\footnote{Note that this uncertainty must be computed by binning over the individual scan measurements per CCD.} at epoch $j$, $\sigma_{\text{ast}}$ is the astrometric jitter term introduced above, and the sum runs over the $N_j$ measurements in the epoch time series.

\section{Orbital Injection Experiments} \label{sec:joint_fitting}

\subsection{Reanalysis of the Sample} \label{sec:refitting}

We benchmarked our modeling pipeline against \citet{Rosenthal2021} by fitting the catalog RVs for all CLS systems hosting a confirmed planet, to ensure internal self-consistency between the RVs and the simulated astrometric dataset. Discrepancies between our best-fit RV solutions and those of \citet{Rosenthal2021} could disrupt the reliability of the joint fitting experiment; therefore, we refitted the sample rather than adopting the catalog posteriors (which also preserves parameter covariances).


The fitting was conducted using only the sum of the RV-only likelihood functions $\mathcal{L}_{\rm RV}^k$, and the basis was reduced to only capture parameters constrained by RVs (i.e., no stellar coordinates, proper motion, parallax, inclination, ascending node, or flux-contamination). In this case, the orbital basis reduced to the same basis used in \citet{Rosenthal2021} aside from our use of the phase $\phi$ instead of $T_p$. For the underlying models, we assumed the same number of Keplerian signals in each system from the CLS catalog (the only exception being detailed in Section \ref{sec:nbody}).
 

An initial guess of the orbital parameters of each companion was taken to be the median value of the posterior reported in the CLS catalog, with two parameters determined separately. We found that the use of the argument of pericenter $\omega$ was inconsistent in that the reported values were ambiguous in their reference to the stellar or planetary orbit (e.g., \citealt{Householder2022}). Therefore, with the other parameters held constant, the parameters $\omega$ and $\phi$ were allowed to float freely. Along with the offset and polynomial terms, initial guesses for these parameters were determined through numerical maximization of the log-likelihood in \texttt{SciPy}. 

Posterior distributions were then estimated using a Markov-Chain Monte Carlo (MCMC) approach implemented with \texttt{emcee} \citep{ForemanMackey2013}. We initialized at least twice as many walkers as free parameters, and allowed them to roam for 80,000 steps with the first 50\% taken to be burn-in. Convergence was assessed through manual inspections of each system's trace plots, and by comparison to previously published solutions and uncertainties. The log-posterior values were saved with the chains to measure the maximum-a-posteriori (MAP) solutions.

\begin{figure*}
    \centering
    \includegraphics[width=1.0\linewidth]{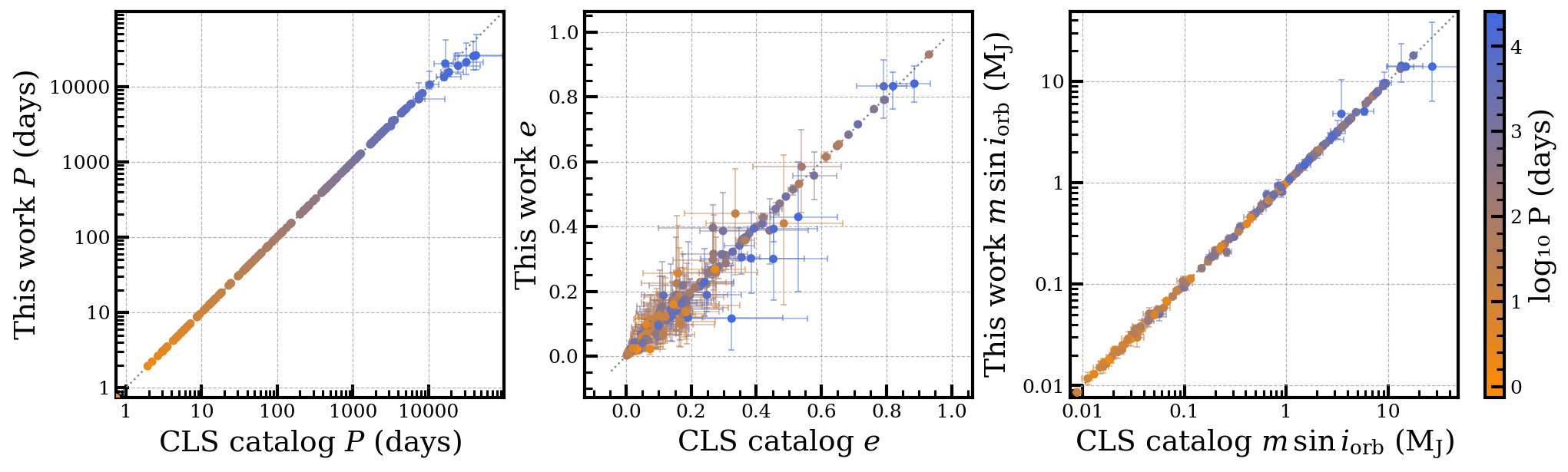}
    \caption{Measured orbital parameters (period, eccentricity, and minimum mass) for all planets in the sample as compared to the original measurements in \citealt{Rosenthal2021}. Every one of the 178 planets is consistent at the $2\sigma$ level across all three parameters.}
    \label{fig:rv_comparison}
\end{figure*}

Every system found orbital solutions consistent to within $2\sigma$ of the CLS catalog.  Figure \ref{fig:rv_comparison} shows a comparison of the posterior distributions of the orbital periods, masses, and eccentricities of all bona fide planets in the sample (although all RV signals, including those owed to systematics or activity, were included in the fitting for consistency). Of the 178 planets considered, only 13 planets differ in $P$, $e$, or $\msini$ by more than $1\sigma$ from the original fits, indicating excellent internal consistency. 

\subsection{Simulation of the Astrometric Datasets} \label{sec:astro_simulation}

We now turn to the primary contribution of this work: assessing the astrometric detectability of the RV-constrained orbits. We constructed a set of astrometric datasets based on the posterior samples from the RV-only fits and then performed joint RV/astrometry fits. To account for the range of orbital configurations allowed by the RV posteriors, the MCMC chains from Section \ref{sec:refitting} were sampled directly, with each iteration proceeding as follows:

\begin{enumerate}
\item Sample an orbital configuration from the MCMC chains. For multi-planet systems, the same chain index was used for all planets to preserve the correlations between their orbital parameters.
\item Uniformly sample the unconstrained astrometric parameters $\cos i_\text{orb}$ and $\Omega$, thereby specifying the true planetary mass(es).
\item Generate a synthetic astrometric dataset for the resulting orbital configuration.
\item Adopt uniform priors on the RV-constrained orbital parameters bounded by the $5\sigma$ intervals from the refitting in Section \ref{sec:refitting}. The stellar mass $M_\star$ retains its independent Gaussian prior.
\item Initialize the joint fit using the MAP RV-posterior parameters, supplemented by a grid search over $\cos i_\text{orb}$ and $\Omega$. We then optimized the full parameter vector $\theta^\text{sys}$ using \texttt{SciPy}.
\item Sample the joint posterior using these priors and the total likelihood $\mathcal{L}_{\text{tot}}$.
\end{enumerate}

\begin{figure}
    \centering
    \includegraphics[width=0.95\linewidth]{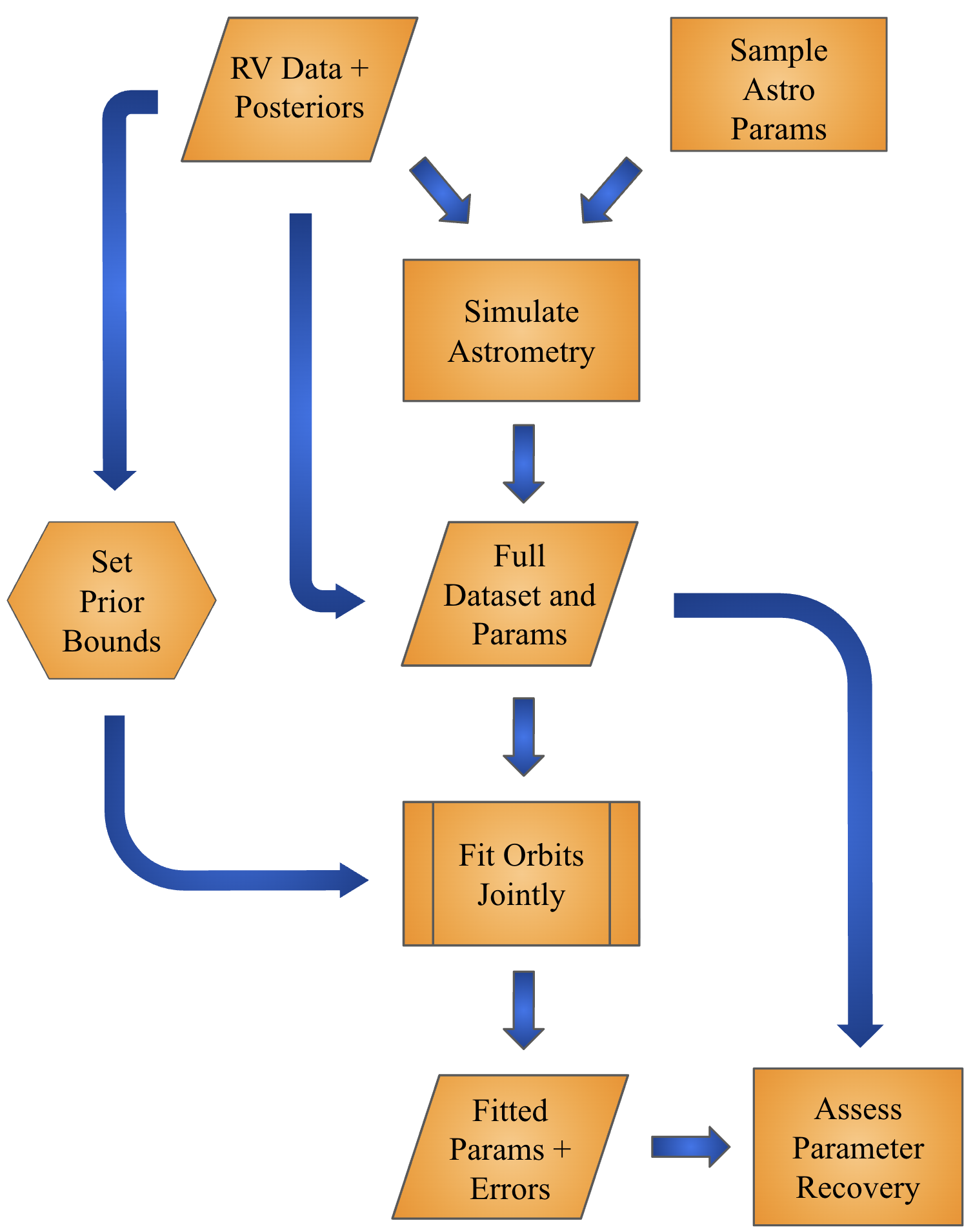}
    \caption{Flowchart for the orbital injection experiment in Section \ref{sec:astro_simulation}. Within each system the simulations are independent, but our predictions are derived from the ensemble of orbits.}
    \label{fig:flowchart}
\end{figure}

\noindent This procedure was repeated for 100 trials per system to account for the unknown values $\cos i_\text{orb}$ and $\Omega$ and other intrinsic uncertainties. A flowchart of the full procedure is shown in Figure \ref{fig:flowchart}. 

Realistic DR4 epoch astrometry was simulated with the \texttt{gaiamock} package \citep{ElBadry2024} using the function \texttt{predictastrometryluminousbinary} with the secondary-to-primary (planet-to-star) flux ratio set to zero. Internally, reflex motion due to companion orbits is translated into celestial coordinates using the Thieles-Innes constants (see Section 4.3 of \citealt{ElBadry2024}). The function outputs a realization of the Gaia observation epochs $\tau_j$ (determined by the stellar coordinates and with a random 10\% of observations discarded) the associated scan angles $\psi_j$, parallax factors $\Pi_j$, and the observed along-scan centroids $\eta_j$ and errors $\sigma_{\eta,j}$. 

For systems with multiple planets, the joint astrometric time-series was simulated by co-adding the independent time-series computed for each of the $N_{c}^{\rm real}$ real planetary companions (that is, excluding activity or instrumental artifacts), and then removing\footnote{Noise realizations in each call of the function were taken to be the same, avoiding addition of noise in quadrature.} the astrometric trajectory of the lone host star $N_c^{\rm real}-1$ times. For convenience, the COM time-series was computed using an additional call of \texttt{predictastrometryluminousbinary} but with the companion mass set to zero (essentially, the five-parameter single-star solution). We note that this method assumes the astrometric signals add linearly, which can be violated in cases of strong planet-planet dynamical interactions, but suffices for the systems considered.

\subsection{Sampling}

The parameters of the joint-fitting basis were then fit to the RV and astrometric datasets simultaneously using \texttt{emcee}. We used the linearization scheme in Appendix \ref{sec:linearization} to reduce convergence time and number of parameters (with a single exception described in the following section). The number of walkers was set to be either 64 or twice the number of free parameters (whichever was larger). Each orbit was sampled for 30,000 steps, and the first 20,000 steps were discarded as burn-in. We again confirmed convergence by manually inspecting the posterior chains.

\subsection{Special Cases: N-body Sampling} \label{sec:nbody}

In most widely spaced planetary systems, orbital periods are not near integer ratios \citep{Wright2011}, allowing the planets to be modeled as independent Keplerian signals. Near mean-motion resonances (MMRs), however, planet--planet interactions perturb the orbits and imprint deviations in the RVs that can bias inferred parameters. These effects are generally modest, but can become significant in systems that are very close to resonance. Therefore, we experimented with modeling the RVs of four CLS systems with multiple giant planets near strong, first-order resonances with a self-consistent N-body model. Those systems are GJ~876, HD~37124, HD~82943, and HD~128311. As discussed in Section~\ref{sec:mutual_inc_results}, these are also promising targets to measure precise mutual inclinations. 

The N-body model specifies the system at a reference epoch and integrates directly over the baseline, converting instantaneous coordinates and velocities at the relevant epochs to RV and astrometric observables. We used \texttt{REBOUND} \citep{Rein2012} with the \texttt{WHFast} symplectic integrator \citep{Rein2015}, adopting a time step of one twentieth of the innermost orbital period. In the RV-only fits, the planets were assumed to share a common orbital inclination \citep[e.g.,][]{Rivera2010}, allowing the RVs to constrain their true masses. Unlike the Keplerian model, orbital elements ($P$, $e$, etc.) evolve throughout the integration rather than remaining fixed at their reference-epoch values. We neglected general relativity and tides, which are negligible over the relevant orbital periods and decades-long baselines. 


For HD~37124, HD~82943, and HD~128311, replacing the Keplerian model with the N-body solver produced little improvement in the RV fits, with $\chi^2$ changes of less than a few percent. The corresponding $\Delta\mathrm{BIC}$ values favored the simpler Keplerian model by 5.1, 21.4, and 14.5 for HD~37124, HD~82943, and HD~128311, respectively. We emphasize that this is not an argument against these planets being in resonance, nor an argument that an N-body fit may not be warranted in the true DR4 data, but rather, that the quality of the RV dataset alone is not sufficient to justify a deviation from a Keplerian approximation for the purposes of this modeling. Moreover, joint RV/astrometry fits (in which coplanarity was relaxed) yielded essentially identical forecasted mutual-inclination precisions, indicating that the N-body treatment provided no additional constraining power from the astrometry. We therefore modeled these three systems with the standard Keplerian pipeline.

The exceptional case was GJ~876, which demonstrates notoriously strong resonant dynamics \citep{Rivera2005,Rivera2010,Nelson2016,Millholland2018}. We compared the Keplerian model, which included three periods attributed to planets and three to stellar activity \citep{Rosenthal2021}, to an N-body model, which contained all four planets, including the outermost planet which was not modeled in CLS (GJ~876~e). The N-body model to the RVs was overwhelmingly favored to the Keplerian model by $\Delta \text{BIC}=886.0$. Moving forward, all results for GJ~876 are presented are with the N-body model\footnote{Given the departure from Keplerians, this was also the sole system where linearized sampling was not used.}, which includes the additional fourth planet (e), bringing the total planet count discussed to 179. 

\begin{figure*}
    \centering
    \includegraphics[width=0.8\linewidth]{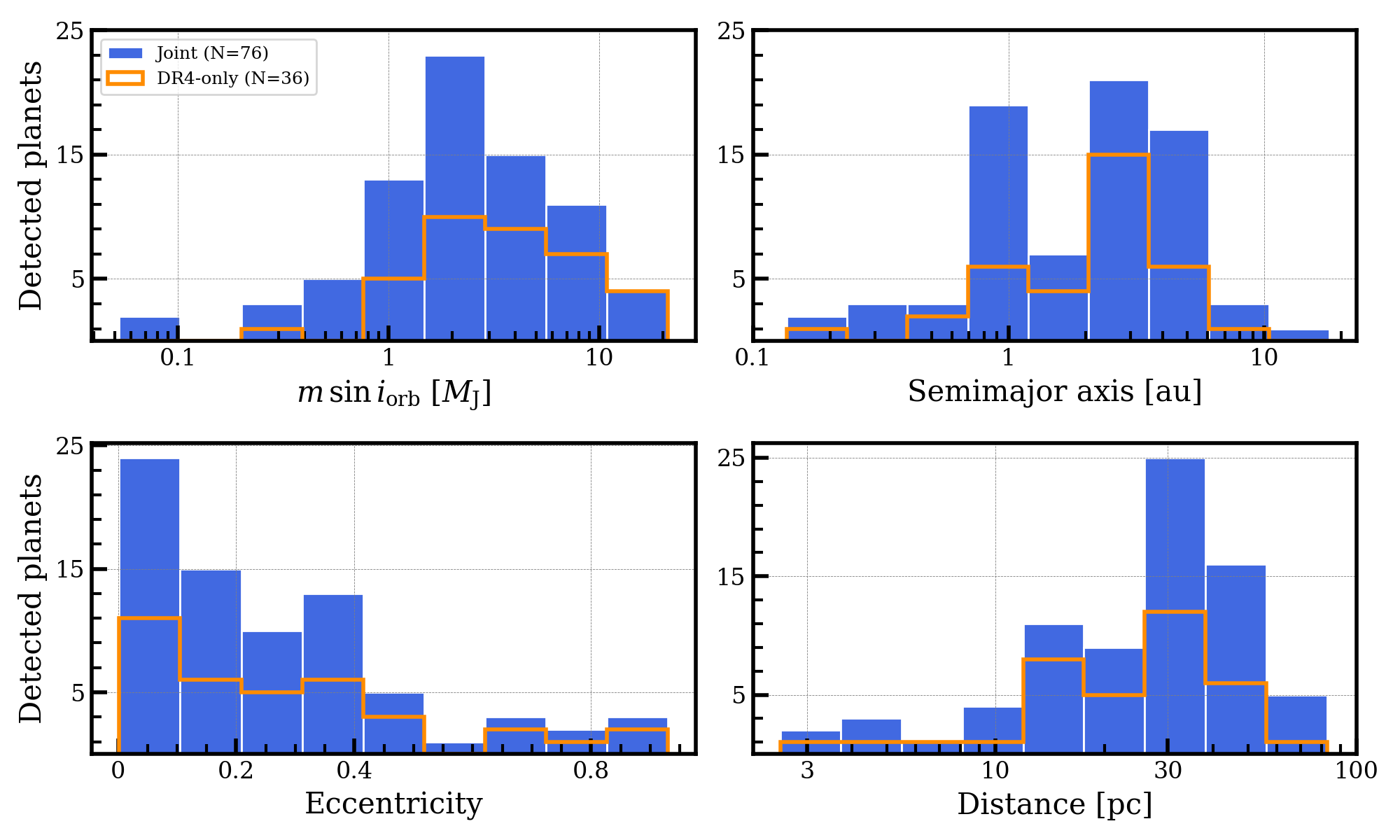}
    \caption{Histograms of the fundamental properties for all planets which permit astrometric detection. For comparison, those detected in the astrometry only (without use of the RVs) are shown in orange.}
    \label{fig:histograms}
\end{figure*}

\subsection{Fits to Astrometry Only}

We executed the injection and recovery experiment twice --- once exactly as described, and once without use of the RV data to assess the role of the RVs in improving the sensitivity of DR4. Similar to our RV refitting in Section \ref{sec:refitting}, we excluded all model parameters which were not relevant (here, RV offsets, jitters, and polynomial terms). We also removed the $5\sigma$ priors to allow the astrometric solution to roam freely.

\section{Results} \label{sec:results}

We now present the results of the RV/astrometry fitting experiments described in Section \ref{sec:joint_fitting}.

\subsection{Forecasted CLS Orbit Catalog}

Fit quality was assessed by estimating the per-orbit $1\sigma$ error on the inclination angle\footnote{\citet{Lammers2026} used fractional errors on orbital period to diagnose fit quality which is generally well known by the RVs, and does not probe the astrometric impact in the joint experiment.} $\Delta i_\text{orb}$ across all the injection tests.  Our criterion for a well-resolved orbit was an error of less than $20^\circ$, which indicates a precision on the true planetary mass of $\lesssim10\%$. The typical poorly-constrained orbit had $\Delta i_\text{orb}>30^\circ$. The orbital parameters and distances of the detected systems are shown in Figure \ref{fig:histograms}. 

\begin{figure}
    \centering
    \includegraphics[width=0.95\linewidth]{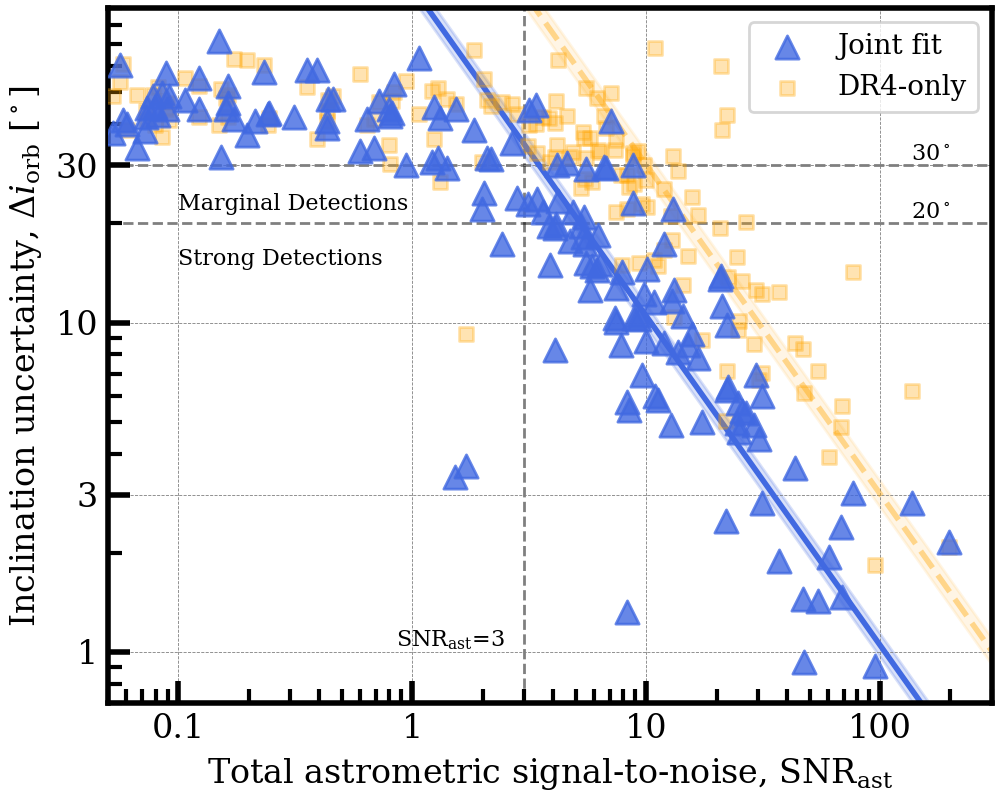}
    \caption{Estimated error on the companion  orbital inclination as a function of total astrometric SNR (Equation \ref{eq:snr}). Our detection threshold of $\Delta i_\text{orb}=20^\circ$ is shown as a dashed line. In the joint model, orbits are marginally ($\Delta i_\text{orb}\approx 30^\circ$) detectable for $\snrast$ above 3, while an astrometric-only model is insensitive until $\snrast\sim$~10. A handful of significant outliers in $\Delta i_\text{orb}$ exist at low $\snrast$, owed to highly eccentric orbits with pericenter passage coinciding with the Gaia epochs (e.g., HD~120066~b).}
    \label{fig:inc_vs_snr}
\end{figure}

In agreement with other works, we found that the quality of the fit was well-described as an inverse function of $\snrast$. To demonstrate this, Figure \ref{fig:inc_vs_snr} shows $\Delta i_\text{orb}$ for all CLS planets as a function of $\snrast$. The joint fit was successful in detecting 76 planets with an error of $\Delta i_\text{orb}<20^\circ$. For comparison, Figure \ref{fig:inc_vs_snr} also shows the astrometry-only runs, which recovered fewer planets by a factor of two (36 planets). 

We fit an inverse function of $\snrast$ to the joint data between $\snrast=1$ and $\snrast=30$, and empirically optimized a drop in sensitivity as a power law in orbital period, giving a power\footnote{This can be understood physically as the curvature contribution to the orbital arc within the time baseline, scaling as $(1/P)^2$.} of -2 in $\mathcal{F}(P)$ (Equation \ref{eq:period_penalty}). That fit indicates a marginal detection of the orbit for values $\snrast\gtrsim3$, and nearly ubiquitous detections for $\snrast\gtrsim7$. When only DR4 data was fit (no RV constraints), strong detections occurred when $\snrast\gtrsim15$. Assuming an approximate sky-averaged $N_j\approx80$ observations per star in DR4, this is comparable to a per-epoch ratio of $\alpha_\textrm{ast}/\sigma_\eta\approx1.6$, which agrees well with \citet{Lammers2026}.

The full list of detected planets is in Table \ref{tab:all_detections}. Among these planets, the typical fractional error on the companion mass was 7\%, the fractional eccentricity error was 10\%, and the fractional period error was 0.15\%. The planets span masses as low as Uranus (GJ~876~e, $m\sim13 M_\oplus$; \citealt{Rivera2010}), semi-major axes as wide as 17 au (HD~120066~b; \citealt{Blunt2019}), and host stars as distant as 83 pc (HD~4203; \citealt{Vogt2022}). Although the joint fits yielded substantially more detected planets than the astrometry-only fits, their properties are broadly similar (Figure \ref{fig:histograms}). A Welch's $t$-test found that joint-fitted planets were, on average, less massive than those in the astrometry-only sample (3.7 versus 5.2 $M_\mathrm{J}$; $p=0.04$), with no other significant differences. We found that incorporating RVs lowers the astrometric amplitude (and hence mass) required to make a detection but does not fundamentally change the planetary architectures that are amenable to astrometric detection in the first place.

\subsection{Detections of Multiple Giant Planets} \label{sec:mutual_inc_results}

\begin{figure}
    \centering
    \includegraphics[width=1.0\linewidth]{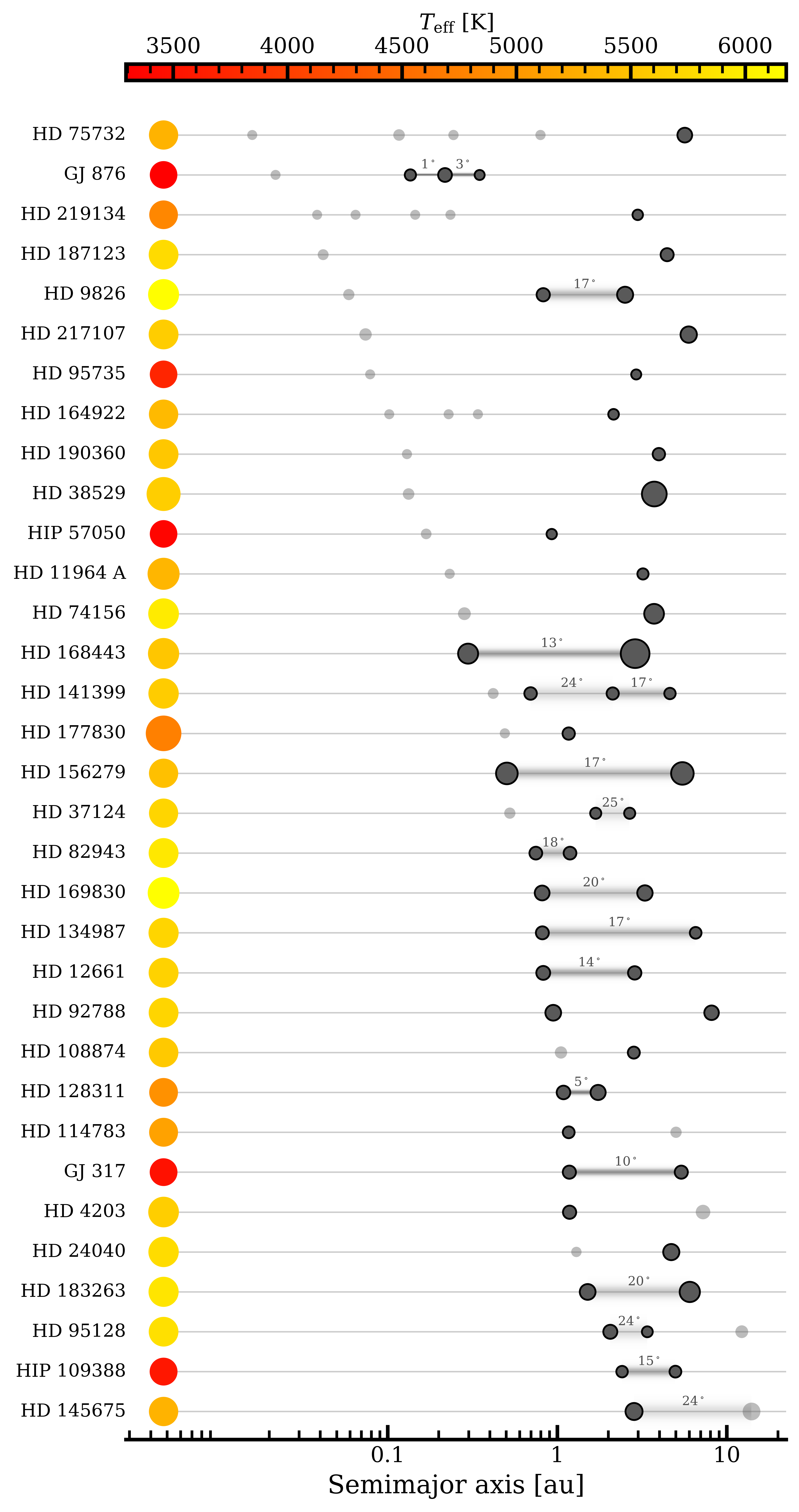}
    \caption{Graphic of all multiplanet CLS systems hosting a detection, with detected planets bolded. Stellar hosts are plotted next to their respective names, with marker sizes scaled by physical radius and colored by $T_\text{eff}$. Planets are plotted according to orbital distance and sized by $m \sin i_\text{orb}$. For systems where the mutual inclinations can be constrained, shaded bands indicate the degree of precision between the planets of interest.}
    \label{fig:all_multiplanets}
\end{figure}

Of the 179 planets considered here, 108 ($\sim60\%$) reside in systems with multiple known planets. This fraction is similar among our forecasted detections, with 51 of 76 detected planets belonging to multiplanet systems. Figure \ref{fig:all_multiplanets} illustrates these detections. In many systems, multiple orbits are resolved with sufficient significance to measure their mutual inclinations. The mutual inclination angle $\Phi$ between the orbits of two planets $b$ and $c$ is given by
\begin{align} \label{eq:mutualinclination}
\cos{\Phi}= & \,\cos{i_{\text{orb},b}}\cos{i_{\text{orb},c}}\\
 &+\sin{i_{\text{orb},b}}\sin{i_{\text{orb},c}}\cos{(\Omega_b-\Omega_c)}.
\end{align}
In cases where the orbital inclinations and longitudes of ascending node are well-determined by the joint fit, the true mutual inclination may be directly constrained (not just the difference between the sky-projected orbital inclinations). 

Using our simulations, we estimated the $1\sigma$ uncertainty $\Delta\Phi$\footnote{Note that the realized value of $\Delta \Phi$ will be lower for systems which are highly inclined. Our experiment was performed agnostic to inclination measurements in the literature.} on the mutual inclination $\Phi$ for all planet pairs in multiplanet systems. As discussed in Section \ref{sec:astro_simulation}, $\cos i_\text{orb}$ and $\Omega$ were sampled from isotropic priors, which induces an isotropic prior on $\Phi$. We therefore exclude from the following analysis any planet pairs for which $\Delta\Phi>25^\circ$, as such large uncertainties indicate a poorly constrained configuration in which $\Phi$ is likely prior dominated. For reference, the isotropic prior in $\cos\Phi$ is similar to a width of $\Delta\Phi\approx 40^\circ$, which motivated our threshold.

We found that the true mutual inclination angle may be constrained in 16 individual planetary systems, summarized in Table \ref{tab:mutual_inc_sample}, which includes $\msini$, $a$, $P$, and $\snrast$ to place the architectures in context with detection significance. All qualifying pairs are also highlighted in Figure \ref{fig:all_multiplanets}. The sample is generally rich in its diversity of properties, ranging from separations of 0.1 to 14 au. When the RVs were neglected and only the astrometry was fit, no mutual inclinations were detected.

The four-planet systems HD~141399 and GJ~876 were unique in containing two detected inner–outer planet pairs. In HD~141399, the (c,~d) and (d,~e) pairs were both detected, while in GJ~876, the (c,~b) and (b,~e) pairs were detected. In each system, the mutual inclination between the innermost and outermost detected planets provides a consistency check on the measured angles. However, because this measurement is geometrically dependent\footnote{Technically, the pairs mentioned are not strictly independent either, as their posteriors can be correlated. However, given the approximate nature of $\Delta\Phi$, we did not attempt to account for this.} on the other two, we exclude those pairs from Table \ref{tab:mutual_inc_sample} and the ensemble statistical analysis (Section \ref{sec:discussion}). Other CLS systems may permit similar high-multiplicity characterization if they have sufficiently low inclinations (e.g., the giant planets in HD~9826, which are thought to have $i_\text{orb}\approx20^\circ$; \citealt{McArthur2010}). In GJ~876, the $\Phi$ values are constrained remarkably well with the N-body model, with $\Delta \Phi\sim 1^\circ$. For comparison, under the Keplerian model (which excluded planet e), the (c,~b) pair of planets were measured with $\Delta \Phi\approx16^\circ$, over ten times less precise.

\begin{deluxetable*}{l cccccc cccccc c}
\tablecaption{Planet parameters and mutual inclination precision for the 16-system sample (18 pairs of planets) with an uncertainty of $\Delta \Phi \le 25^\circ$. The pairwise measurements in the HD~141399 and GJ~876 systems are marked with an asterisk ($^*$). \label{tab:mutual_inc_sample}}
\tablehead{
& \multicolumn{6}{c}{Planet 1} & \multicolumn{6}{c}{Planet 2} & \\
\cline{2-7} \cline{8-13}
\colhead{Star} & \colhead{Index} & \colhead{$m\sin i_\text{orb}$} & \colhead{$a$} & \colhead{$P$} & \colhead{$\snrast$} & \colhead{$e$} & \colhead{Index} & \colhead{$m\sin i_\text{orb}$} & \colhead{$a$} & \colhead{$P$} & \colhead{$\snrast$} & \colhead{$e$} & \colhead{$\Delta \Phi$} \\
& \colhead{} & \colhead{($M_{\rm J}$)} & \colhead{(au)} & \colhead{(d)} & \colhead{} & \colhead{} & \colhead{} & \colhead{($M_{\rm J}$)} & \colhead{(au)} & \colhead{(d)} & \colhead{} & \colhead{} & \colhead{(deg)}
}
\startdata
GJ 876$^*$ & c & 0.7 & 0.1 & 30 & 9.2 & 0.26 & b & 2.1 & 0.2 & 61 & 48.2 & 0.03 & 1 \\
GJ 876$^*$ & b & 2.1 & 0.2 & 61 & 48.2 & 0.03 & e & 0.04 & 0.3 & 124 & 1.5 & 0.10 & 3 \\
HD 128311 & b & 2.0 & 1.1 & 453 & 24.6 & 0.27 & c & 3.2 & 1.7 & 917 & 60.1 & 0.18 & 5 \\
GJ 317 & b & 1.9 & 1.2 & 696 & 67.5 & 0.10 & c & 1.7 & 5.4 & 6784 & 22.3 & 0.19 & 10 \\
HD 168443 & b & 7.9 & 0.3 & 58 & 8.7 & 0.53 & c & 17.9 & 2.9 & 1750 & 198.8 & 0.21 & 13 \\
HD 12661 & b & 2.3 & 0.8 & 262 & 7.4 & 0.36 & c & 1.9 & 2.9 & 1696 & 22.4 & 0.02 & 14 \\
HIP 109388 & b & 0.9 & 2.4 & 1938 & 72.5 & 0.04 & c & 1.1 & 5.0 & 5733 & 20.9 & 0.09 & 15 \\
HD 9826 & c & 2.0 & 0.8 & 241 & 5.9 & 0.27 & d & 4.1 & 2.5 & 1282 & 37.7 & 0.31 & 17 \\
HD 141399$^*$ & d & 1.2 & 2.1 & 1074 & 10.9 & 0.04 & e & 0.8 & 4.6 & 3445 & 4.0 & 0.08 & 17 \\
HD 156279 & b & 9.6 & 0.5 & 133 & 17.1 & 0.65 & c & 9.5 & 5.5 & 4763 & 30.7 & 0.26 & 17 \\
HD 134987 & b & 1.6 & 0.8 & 258 & 9.7 & 0.23 & c & 0.9 & 6.5 & 5854 & 4.1 & 0.15 & 17 \\
HD 82943 & c & 1.6 & 0.7 & 220 & 6.3 & 0.43 & b & 1.6 & 1.2 & 442 & 9.9 & 0.16 & 18 \\
HD 169830 & b & 2.9 & 0.8 & 226 & 5.4 & 0.31 & c & 3.6 & 3.3 & 1837 & 24.6 & 0.26 & 20 \\
HD 183263 & b & 3.7 & 1.5 & 625 & 14.0 & 0.38 & c & 8.0 & 6.0 & 5017 & 15.3 & 0.06 & 20 \\
HD 141399$^*$ & c & 1.3 & 0.7 & 202 & 4.1 & 0.05 & d & 1.2 & 2.1 & 1074 & 10.9 & 0.04 & 24 \\
HD 95128 & b & 2.4 & 2.1 & 1077 & 29.9 & 0.01 & c & 0.5 & 3.4 & 2290 & 6.3 & 0.17 & 24 \\
HD 145675 & b & 4.9 & 2.8 & 1766 & 142.5 & 0.37 & c & 5.2 & 14.0 & 19372 & 6.9 & 0.33 & 24 \\
HD 37124 & c & 0.6 & 1.7 & 887 & 5.7 & 0.10 & d & 0.6 & 2.7 & 1766 & 9.0 & 0.14 & 25 \\
\enddata
\end{deluxetable*}


\subsection{Mass Upper Limits from Non-Detections}

Precise measurements of orbital angles are not the only use for the Gaia astrometry. Even in the absence of a significant astrometric detection, the data can rule out nearly face-on orbits ($\sin i_\mathrm{orb}\rightarrow0$), thereby placing upper limits on companion masses. These constraints can help validate the planetary nature of RV candidates and reduce the mass range that must be marginalized over when inferring population properties such as the planetary mass function \citep{VanZandt2026}. There are approximately 16 planets for which a significant astrometric detection is unlikely, but the inclination can still be constrained to $\Delta i_\mathrm{orb}\sim20^\circ$--$30^\circ$ (Figure \ref{fig:inc_vs_snr}). Thus, even sub-threshold astrometric signals can meaningfully exclude brown-dwarf companions and refine the planetary mass function across the sample.

\begin{figure}
    \centering
    \includegraphics[width=0.9\linewidth]{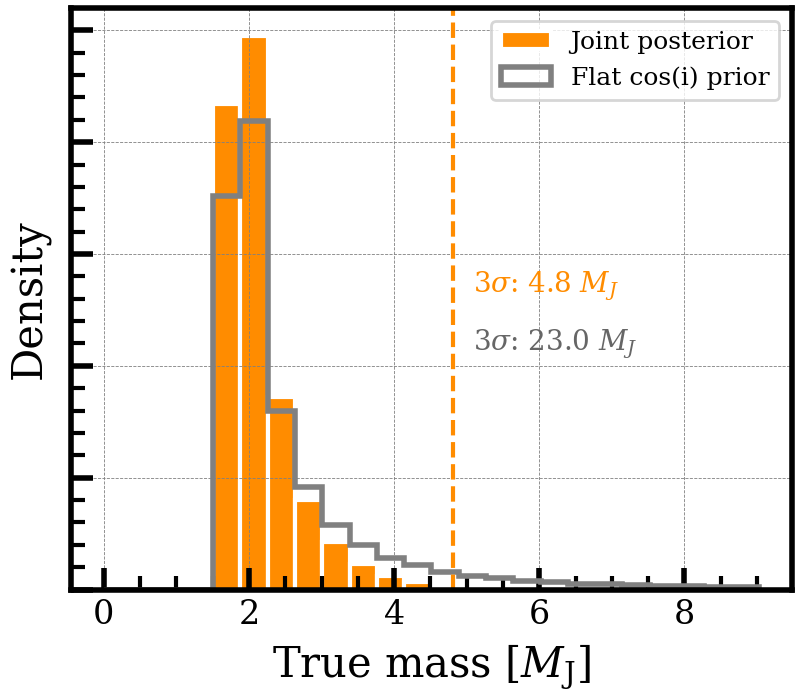}
    \caption{Posterior distributions of a joint model for the planet HD~13931 ($m \sin i_\text{orb}\approx1.9M_\text{J}$, $P\approx12$~yr) with an injected $\cos i_\text{orb}=0$. This realization had $\snrast\sim2$, so the planet is not strongly detected, but the fit rules out face-on orientations and refines the upper limit on the true mass. This run had $\Delta i_\text{orb}\sim30^\circ$, which would be considered a marginal to non-detection in the ensemble.}
    \label{fig:no_face_on}
\end{figure}

\section{Discussion} \label{sec:discussion}


Mutual inclinations have emerged as a key probe of exoplanet architectures, as the field shifts from asking which planets exist (occurrence) to how planets and their environments are arranged relative to one another. Recent work, often enabled by Gaia astrometry, has constrained the inclinations between stars and their protoplanetary disks \citep{Biddle2025,Biddle2026}, inner transiting planets and outer giants \citep{Xiao2026b}, transiting planets and stellar binary orbits \citep{Zhang2026}, and planetary spins and orbits \citep{Poon2025,Lammers2026b}. However, these techniques typically yield only minimum mutual inclinations, i.e., the difference between sky-plane inclinations because the longitudes of the ascending nodes remain unknown. The CLS-Gaia project will instead deliver true orbit--orbit mutual inclinations for a population of giant-planet systems. 

In the following sections, we test the statistical power of the sample. In Section \ref{sec:phi_models}, we address a qualitative question --- can we infer whether the mutual inclinations are aligned, broadly distributed, or a mixture model of aligned and uniformly distributed systems? In Section \ref{sec:phi_recovery}, we perform a distinct quantitative experiment, where we predict how well the model hyperparameters could be estimated.

\subsection{Model Selection} \label{sec:phi_models}

\begin{figure*}
    \centering
    \includegraphics[width=0.8\linewidth]{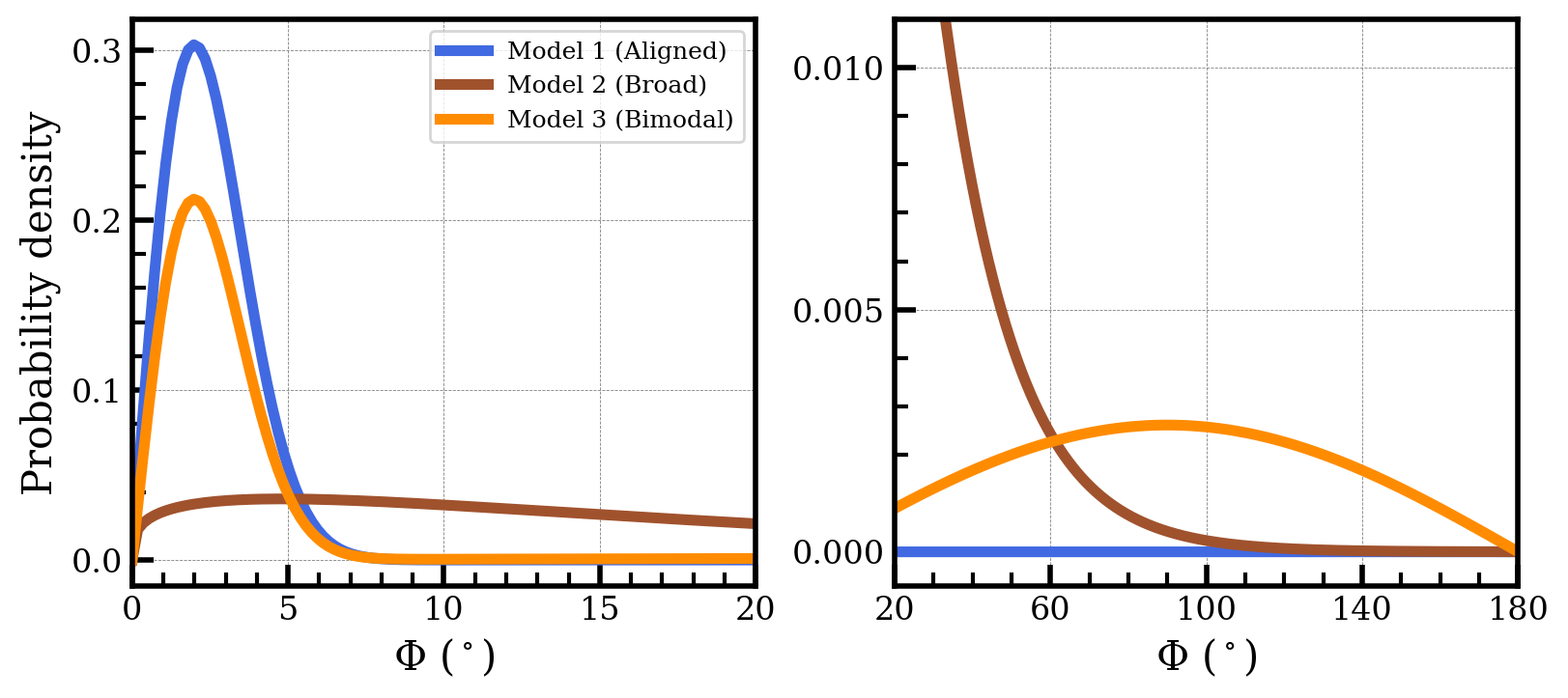}
    \caption{Probability densities of each toy model for the mutual inclination ($\Phi$) distribution in Section \ref{sec:phi_models}. The regime below (left) and above (right) $\Phi=20^\circ$ are differentiated for more appropriate scaling. Moderate mutual inclinations ($10^\circ-30^\circ$) provide the most support for Model 2, while the presence of retrograde orientations most directly implicate Model 3. The absence of both suggests a tight distribution near alignment, consistent with Model 1.}
    \label{fig:three_toy_models}
\end{figure*}

We tested three disparate forms that the mutual inclination distribution might take, with a physical interpretation attached to each. Those models were:
\vspace{\baselineskip}

\noindent\textbf{Model 1 --- Single Aligned Mode/Compact-Multi-Like}: Mutual inclinations distributed similarly to those inferred for Kepler-like compact, multi-transiting systems. We adopt a single Rayleigh distribution with scale parameter $\sigma_\Phi=2^\circ$ \citep{Fabrycky2014,Fang2012,Lissauer2011}, with probability density given by:

\begin{equation}
    p_1(\Phi \mid \sigma_\Phi) = \frac{\Phi}{\sigma_\Phi^2} \exp\!\left(-\frac{\Phi^2}{2\sigma_\Phi^2}\right)
\end{equation}
This model represents a dynamically cold architecture, consistent with the near-coplanarity expected from assembly within a common protoplanetary disk.
\vspace{\baselineskip}
    
\noindent\textbf{Model 2 --- Broad Distribution/Eccentricity-Like}: Motivated by the equipartition of angular momentum deficit (AMD), the mutual inclinations of giant planets follow a distribution analogous to their eccentricities. We adopt a Gamma distribution, which provides a good description of the observed eccentricity distribution measured by RVs \citep{Stevenson2025,Kipping2013,Wang2011,Juric2008}. The probability density is given by
\begin{equation}
    p_2(\Phi \mid \alpha_\Phi, \beta_\Phi) = \frac{\Phi^{\alpha_\Phi - 1} e^{-\Phi/\beta_\Phi}}{\Gamma(\alpha_\Phi)\,\beta_\Phi^{\alpha_\Phi}}
\end{equation}
where we set the shape parameter $\alpha_\Phi=1.3$ and the characteristic scale $\beta_\Phi =16^\circ$. These parameters are scaled from their eccentric counterparts by a 1:1 equipartition assumption of $\langle\Phi^2\rangle\approx2\langle e^2\rangle$ (see Appendix \ref{sec:equipartition}). This model represents an excited population in which mutual inclinations are more broadly distributed, analogous to the eccentricity excitation driven by thermalizing processes such as scattering. 
\vspace{\baselineskip}

\noindent\textbf{Model 3 --- Distinct Dynamical Modes/Stellar Obliquity-Like}: A bimodal distribution comprising distinct dynamically cold and hot subpopulations, analogous to the distribution inferred for stellar obliquities of transiting giant planets, typically orbiting at scales of $\approx0.1$~au \citep{Knudstrup2024,Siegel2023,Campante2016,Winn2009,Fabrycky2009}. With frequency $f=0.7$, $\Phi$ is drawn from an aligned Rayleigh distribution with scale $\sigma_\Phi=2^\circ$; with frequency $1-f=0.3$, mutual inclinations are drawn isotropically. The density is then
\begin{equation}
\begin{aligned}
    p_3(\Phi \mid f, \sigma_\Phi) = {} &
    f \cdot \frac{\Phi}{\sigma_\Phi^2}
    \exp\!\left(-\frac{\Phi^2}{2\sigma_\Phi^2}\right) \\
    & + (1-f) \cdot \frac{1}{2}\sin\Phi,
    \qquad \Phi \in [0,\pi].
\end{aligned}
\end{equation}
with $\Phi$ and $\sigma_\Phi$ expressed in radians. This model represents a population in which most systems remain nearly coplanar, while a substantial minority undergo strong orbital evolution that erases the memory of their initial reference plane. The isotropic systems would be the likely precursors for long-range migration and destabilization of inner planetary systems \citep[e.g.,][]{Becker2026}, assuming a connection between the transiting and RV-detected giant planets.
\vspace{\baselineskip}
    
Together, these models span possibilities for the dynamical architectures of giant-planet systems. For intuition, we show the probability densities of each model in Figure \ref{fig:three_toy_models}. 

We tested whether CLS-Gaia data will be able to distinguish between the three models, with the hyperparameters fixed at their fiducial values.  The process involved drawing realizations of the data for each model, computing simulated sets of orbital measurements for the planetary systems, and evaluating the model likelihoods within a hierarchical Bayesian framework. Accounting for the measurement uncertainties and the constraint that the measured mutual inclination must be on the interval [0,$\pi$], the likelihood that the star $s$ is drawn from Model $j$ is given by
\begin{align}
    \mathcal{L}_s(\theta) = \int_0^{\pi} p_j(\Phi \mid &\theta)\,\Big[\mathcal{N}(\Phi_{\mathrm{obs},s} - \Phi;\,0,\Delta \Phi_s^2) \\
    & + \mathcal{N}(-\Phi_{\mathrm{obs},s} - \Phi;\,0,\Delta \Phi_s^2) \\
    & + \mathcal{N}(2\pi - \Phi_{\mathrm{obs},s} - \Phi;\,0,\Delta \Phi_s^2)\Big]\, d\Phi \notag.
\end{align}
The total log-likelihood of the model $\text{ln}\, \mathcal{L}$ was computed by summing $\text{ln}\, \mathcal{L}_s$ over the sample. Because the parameters of each distribution are here held fixed to their fiducial values, the model evidence depends only on the likelihood given the data (i.e., the number of parameters is zero). We considered a model to be disfavored when the evidence difference exceeded $-2 \Delta \text{ln}\,\mathcal{L}=\Delta\mathrm{BIC}=6$, corresponding to a strong preference for the lower-BIC model.

\begin{figure}
    \centering
    \includegraphics[width=1.0\linewidth]{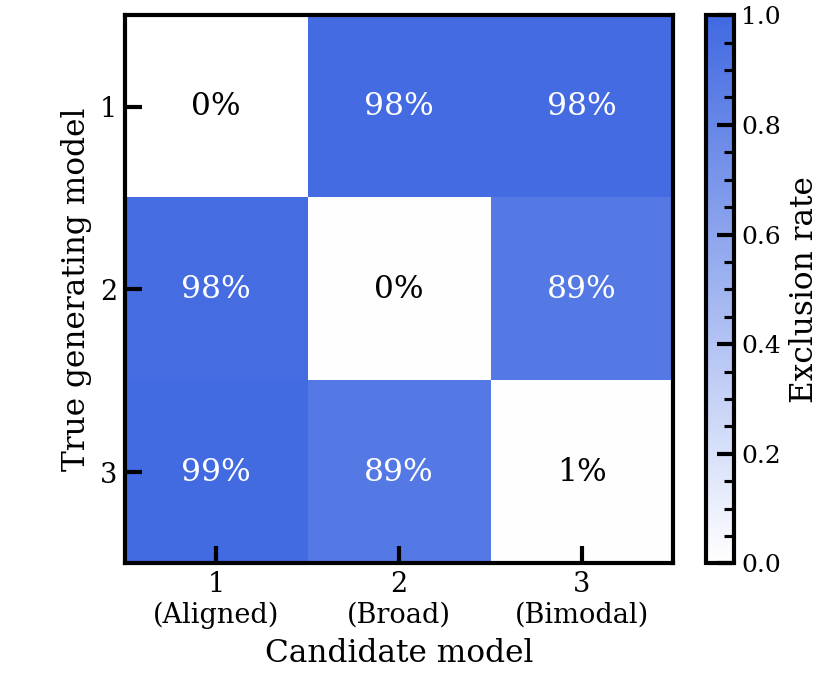}
    \caption{Matrix indicating which model was injected into the sample (rows) and which models were strongly excluded by the data (columns). A value near 100\% indicates a strong constraint against the physical description implied by the model in that column. We again remark that the hyperparameters were held fixed in this experiment to prevent the flexible distributions (particularly Model 1 and Model 2) from changing qualitatively.}
    \label{fig:exclusion_matrix}
\end{figure}

Using this threshold, we tested whether our mutual inclination measurements distinguish between the three distributions. Figure \ref{fig:exclusion_matrix} summarizes 1000 injection tests in which the three models had their likelihood evaluated. In the vast majority of realizations ($\gtrsim90\%$), the data were sufficient to rule out the incorrect models, regardless of what population was injected. The aligned distribution (Model 1) was the most orthogonal of the set, being correctly identified or excluded in nearly 100\% of the realizations. Across all tests, the true model was improperly excluded in $<1\%$ of realizations, implying our likelihood threshold was reliable. Altogether, the CLS-Gaia dataset is likely to suggest a physical picture for where mutual inclinations between giant planets may originate.

\subsection{Parameter Recovery} \label{sec:phi_recovery}

We also explored how well the parameters of the population models that describe mutual inclinations can be determined. To do this, we varied the parameters over a grid of ten injected values, and fit the models using maximum likelihood. For Models 2 and 3, while one parameter was varied, the other was held at the nominal value from Section \ref{sec:phi_models}. At each grid point, we performed 20 realizations to constrain possible systematic biases in the best-fit parameters, and noted the underlying uncertainty per-realization. The threshold for a bias to be significant was if the mean recovered parameter over the grid deviated from the injected value by more than the per-realization measurement uncertainty.

The bias and precision range on each parameter is given in Table \ref{tab:mutual_inc_precision_summary}. We highlight that the scale of the mutual inclination spread (either $\sigma_\Phi$ or $\beta_\Phi$) is likely to be constrained at a precision of a few degrees, allowing a meaningful comparison between the CLS-Gaia sample, the multi-transiting planet sample \citep{Xie2016}, and the transiting planets with hierarchical outer giants sample \citep{Xiao2026b}. We also find it unlikely for any of these model parameters to be strongly biased in the physically plausible ranges we tested.

\begin{deluxetable*}{l c c c c c}
\tablecaption{Parameter recovery: bias and precision for each benchmark model.  \label{tab:mutual_inc_precision_summary}}
\tablehead{
\colhead{Model} &
\colhead{Parameter} &
\colhead{Units} &
\colhead{Biased?} &
\colhead{\shortstack{Injected\\Parameter Range}} &
\colhead{\shortstack{Recovered\\Parameter Precision}}
}
\startdata
Model 1 (Rayleigh) & $\sigma_\Phi$ & deg & No & 2--30 & 1--5 \\
Model 2 (Gamma) & $\alpha_\Phi$ & \nodata & No & 1--3 & 0.5--0.5 \\
Model 2 (Gamma) & $\beta_\Phi$ & deg & No & 5--30 & 2--7 \\
Model 3 (Rayleigh+Iso) & $f$ & \nodata & No & 0.4--0.9 & 0.13--0.08 \\
Model 3 (Rayleigh+Iso) & $\sigma_\Phi$ & deg & No & 2--30 & 1--10 \\
\enddata
\end{deluxetable*}

\section{Conclusion} \label{sec:conclusion}

This work presents the CLS-Gaia project, in which a benchmark sample of nearly 200 planets with decades of RV observations will be combined with epoch astrometry from the Gaia mission. We motivated the use of RVs and astrometry in a joint fit, and described a model by which to do so. Using the CLS RVs, we refit the CLS sample, used those posterior chains to generate mock DR4 astrometry, and fit both datasets over an ensemble of orbits for each planetary system. From these orbits, we made predictions about the fundamental properties of the detectable planets. The important conclusions of this work are summarized below:

\begin{itemize}
    \item We found that 76 planets around CLS stars will benefit from resolved orbits when the Gaia astrometry is fit simultaneously with the RVs. With our threshold of an inclination angle uncertainty of $\Delta i_\text{orb}<20^\circ$, these planets will have typical fraction mass errors of 7\%, fractional eccentricity errors of 10\%, and fractional period errors of 0.15\%. Most of these planets (51/76) were in multi-planet systems. We noted an additional 16 planets where an upper-mass limit can be established in the event of a non-detection. With the DR4 astrometry alone, fewer than half as many planets (36) are likely to be detected. These planets are of noticeably higher mass than the joint-fit sample, highlighting the significance of the RVs in constraining orbital solutions.
    \item When jointly fit, we identified 18 pairs of giant planets across 16 systems that are amenable to meaningful mutual-inclination constraints. Some systems, specifically HD~141399 and GJ~876, additionally allow multiple mutual inclinations to be measured within the same system. More detections may be possible for low inclination (plane-of-sky) systems. Through simulations of toy models for the underlying population, we demonstrated that our sample is sensitive to the shape and width of the mutual-inclination distribution, which can be constrained to within a few degrees. These results establish planet-planet misalignments among RV-detected giant planets as a population-level observable, placing them on a comparable footing with other benchmark populations of orbital angles in the literature.
\end{itemize}

\begin{acknowledgments}

The authors are grateful for discussions with Kareem El-Badry and Konstantin Batygin.  The CLS-Gaia project was conceived in the course Ay 198 at the California Institute of Technology; the authors are grateful for the latitude in course design provided by the Institute. L.B.H.\ and A.S.\ acknowledge support from the National Science Foundation through the Graduate Research Fellowship Program under Grant No. 2139433. This research made use of the NASA Exoplanet Archive, which is operated by the California Institute of Technology, under contract with the National Aeronautics and Space Administration under the Exoplanet Exploration Program.

Some of the data discussed herein were obtained at the W.\,M.~Keck Observatory, which is operated as a scientific partnership among the California Institute of Technology, the University of California, and the National Aeronautics and Space Administration. We wish to recognize and acknowledge the very significant cultural role and reverence that the summit of Maunakea has always had within the indigenous Hawaiian community. We are most fortunate to have the opportunity to conduct observations from this mountain.

\end{acknowledgments}

\facilities{Keck I (HIRES), Automated Planet Finder (Levy), Lick (Hamilton), Exoplanet Archive}

\software{
    \texttt{astropy} \citep{astropy:2013,astropy:2018,astropy:2022},
    \texttt{astroquery} \citep{astroquery},
    \texttt{emcee} \citep{ForemanMackey2013},
    \texttt{gaiamock} \citep{ElBadry2024},
    \texttt{Matplotlib} \citep{matplotlib},
    \texttt{NumPy} \citep{numpy},
    \texttt{pandas} \citep{pandas},
    \texttt{REBOUND} \citep{Rein2012,Rein2015},
    \texttt{SciPy} \citep{scipy}.
}

\appendix

\section{All Detected Planets and Their Parameter Uncertainties}

Contained here is the full sample of detected planets, their properties, and our predicted precision of the joint fit on those properties in Table \ref{tab:all_detections}. As used elsewhere in this work, $\Delta X$ is the empirically-calibrated $1\sigma$ uncertainty level on the parameter $X$. For mass, eccentricity, and period, these uncertainties are normalized by their median values and are stated as a percentage. Note that the eccentricity and period uncertainties are mostly constrained by the RVs and are already well known but are included for completeness.

\startlongtable
\begin{deluxetable*}{l c c c c c c c c c}
\tablecaption{Precision of the 76 detected planets ($\Delta i_\text{orb} \le 20^\circ$) in the joint RV/astrometry fit. \label{tab:all_detections}}
\tablehead{
\colhead{Planet} & \colhead{$\snrast$} & \colhead{$\Delta i_\text{orb}$ (deg)} & \colhead{$m\sin i_\text{orb}$ ($M_J$)} & \colhead{$\Delta m/m$ (\%)} & \colhead{$e$} & \colhead{$\Delta e/e$ (\%)} & \colhead{$P$ (d)} & \colhead{$\Delta P/P$ (\%)} & \colhead{$d$ (pc)}
}
\startdata
HD 168443 c & 198.8 & 2.2 & 18.0 & 2.2 & 0.21 & 0.62 & 1750 & 0.01 & 40 \\
HD 145675 b & 142.5 & 2.8 & 4.9 & 2.1 & 0.37 & 1.2 & 1766 & 0.05 & 18 \\
HD 33636 b & 94.0 & 0.9 & 9.1 & 0.65 & 0.49 & 0.72 & 2109 & 0.05 & 30 \\
HIP 109388 b & 72.5 & 3.0 & 0.9 & 3.4 & 0.04 & 48 & 1938 & 0.5 & 8.8 \\
HD 38529 c & 67.8 & 1.5 & 13.3 & 1.4 & 0.35 & 1.3 & 2133 & 0.08 & 43 \\
GJ 317 b & 67.5 & 2.4 & 1.9 & 2.6 & 0.10 & 21 & 696 & 0.09 & 15 \\
HD 128311 c & 60.1 & 1.9 & 3.2 & 2.5 & 0.18 & 24 & 917 & 0.2 & 16 \\
HD 10697 b & 56.0 & 1.4 & 6.5 & 1.4 & 0.10 & 6.5 & 1076 & 0.05 & 33 \\
GJ 876 b & 48.2 & 0.9 & 2.1 & 2 & 0.03 & 3.7 & 61 & 0.007 & 4.7 \\
HIP 22627 b & 46.3 & 1.4 & 0.8 & 3.3 & 0.16 & 28 & 2322 & 1 & 12 \\
HD 50554 b & 42.5 & 3.6 & 4.3 & 3.2 & 0.45 & 4.2 & 1223 & 0.1 & 31 \\
HD 9826 d & 37.7 & 1.9 & 4.1 & 2.7 & 0.31 & 5.9 & 1282 & 0.07 & 13 \\
HD 156279 c & 30.7 & 6.0 & 9.5 & 3.9 & 0.26 & 1.8 & 4763 & 0.6 & 36 \\
HD 222582 b & 30.3 & 2.8 & 7.9 & 3.1 & 0.76 & 0.64 & 573 & 0.01 & 43 \\
HD 74156 c & 30.1 & 4.4 & 7.6 & 3 & 0.37 & 1.5 & 2453 & 0.2 & 58 \\
HD 95128 b & 29.9 & 6.9 & 2.4 & 5.3 & 0.01 & 74 & 1077 & 0.1 & 14 \\
HD 22049 b & 28.1 & 4.9 & 0.6 & 6.1 & 0.04 & 1e+02 & 2656 & 0.6 & 3.2 \\
HD 87883 b & 28.0 & 5.3 & 2.3 & 3.8 & 0.71 & 1.1 & 3386 & 2 & 18 \\
HD 75732 d & 25.2 & 5.4 & 3.3 & 6.8 & 0.14 & 13 & 4933 & 1 & 13 \\
HD 186427 b & 24.9 & 4.6 & 1.8 & 3.3 & 0.68 & 0.48 & 799 & 0.02 & 21 \\
HD 128311 b & 24.6 & 4.9 & 2.0 & 7.4 & 0.27 & 15 & 453 & 0.1 & 16 \\
HD 169830 c & 24.6 & 5.7 & 3.5 & 5.2 & 0.26 & 5.6 & 1837 & 0.4 & 38 \\
HD 12661 c & 22.4 & 6.3 & 1.9 & 6.3 & 0.02 & 67 & 1696 & 0.1 & 37 \\
GJ 317 c & 22.3 & 13.6 & 1.7 & 11 & 0.19 & 27 & 6784 & 6 & 15 \\
HD 28185 b & 22.2 & 6.4 & 6.0 & 4.4 & 0.07 & 11 & 386 & 0.05 & 39 \\
HD 217107 c & 21.7 & 9.8 & 4.4 & 3.3 & 0.39 & 1.7 & 5139 & 0.3 & 20 \\
HD 181234 b & 21.5 & 2.5 & 9.4 & 2.5 & 0.84 & 4.1 & 7532 & 1 & 47 \\
HIP 109388 c & 20.9 & 13.9 & 1.1 & 10 & 0.09 & 41 & 5733 & 3 & 8.7 \\
HD 190360 b & 20.3 & 11.2 & 1.5 & 4 & 0.32 & 3.7 & 2894 & 0.4 & 16 \\
HD 156279 b & 17.1 & 5.0 & 9.6 & 3.3 & 0.65 & 0.12 & 133 & 0.003 & 36 \\
HIP 83043 b & 16.9 & 7.8 & 0.3 & 10 & 0.20 & 40 & 602 & 0.6 & 10 \\
HD 183263 c & 15.3 & 9.2 & 8.0 & 5 & 0.06 & 11 & 5017 & 0.6 & 55 \\
HD 117207 b & 14.9 & 8.6 & 1.9 & 5.5 & 0.14 & 18 & 2618 & 0.6 & 32 \\
HD 183263 b & 14.0 & 10.4 & 3.7 & 11 & 0.38 & 1.1 & 625 & 0.02 & 54 \\
HD 24040 b & 13.6 & 8.2 & 4.1 & 4.3 & 0.02 & 82 & 3546 & 0.4 & 47 \\
HD 117176 b & 13.0 & 12.6 & 7.3 & 9.2 & 0.40 & 0.33 & 117 & 0.0006 & 18 \\
HD 92788 b & 12.5 & 4.9 & 3.5 & 4.6 & 0.36 & 2 & 326 & 0.01 & 35 \\
HD 219134 h & 12.4 & 11.6 & 0.3 & 17 & 0.04 & 73 & 2118 & 4 & 6.6 \\
HD 210277 b & 12.0 & 8.7 & 1.2 & 15 & 0.47 & 1.2 & 443 & 0.02 & 21 \\
HD 95735 c & 11.6 & 17.3 & 0.05 & 12 & 0.12 & 6.3e+02 & 2917 & 20 & 2.5 \\
HD 40979 b & 11.0 & 5.8 & 3.8 & 5.2 & 0.26 & 8.1 & 265 & 0.06 & 35 \\
HD 141399 d & 10.9 & 11.6 & 1.3 & 12 & 0.04 & 26 & 1074 & 0.3 & 37 \\
HD 154345 b & 10.7 & 6.0 & 0.9 & 8.3 & 0.02 & 1e+02 & 3459 & 2 & 18 \\
HIP 57050 c & 10.3 & 14.6 & 0.2 & 16 & 0.41 & 14 & 533 & 0.7 & 11 \\
HD 82943 b & 9.9 & 8.8 & 1.6 & 9.2 & 0.16 & 28 & 442 & 0.1 & 27 \\
HD 108874 c & 9.7 & 12.2 & 1.1 & 11 & 0.27 & 9.5 & 1743 & 0.6 & 60 \\
HD 134987 b & 9.7 & 6.9 & 1.6 & 6.7 & 0.23 & 3.8 & 258 & 0.02 & 26 \\
HD 114783 b & 9.5 & 10.3 & 1.0 & 6.8 & 0.11 & 18 & 492 & 0.07 & 21 \\
GJ 876 c & 9.2 & 1.3 & 0.7 & 2.4 & 0.26 & 0.75 & 30 & 0.03 & 4.7 \\
HD 37124 d & 9.0 & 10.6 & 0.6 & 14 & 0.14 & 59 & 1766 & 0.6 & 32 \\
HD 168443 b & 8.7 & 10.3 & 8.0 & 5.5 & 0.53 & 0.13 & 58 & 0.0002 & 39 \\
HD 11964 A b & 8.4 & 5.4 & 0.6 & 5.7 & 0.10 & 45 & 1940 & 0.5 & 33 \\
HD 164922 b & 8.2 & 5.8 & 0.3 & 6.1 & 0.05 & 57 & 1197 & 0.3 & 22 \\
HD 114729 b & 7.9 & 14.2 & 0.9 & 8 & 0.10 & 65 & 1134 & 0.5 & 38 \\
HD 178911 B b & 7.8 & 8.6 & 7.1 & 8.3 & 0.11 & 2.2 & 71 & 0.003 & 41 \\
HD 45350 b & 7.5 & 10.3 & 1.8 & 8.8 & 0.79 & 1.2 & 978 & 0.03 & 47 \\
HD 12661 b & 7.4 & 12.8 & 2.3 & 8.9 & 0.36 & 1.5 & 262 & 0.007 & 38 \\
HD 187123 c & 7.4 & 10.0 & 1.8 & 16 & 0.22 & 8.2 & 3412 & 0.8 & 46 \\
HD 80606 b & 6.3 & 14.5 & 4.2 & 14 & 0.93 & 0.18 & 111 & 0.0003 & 64 \\
HD 95128 c & 6.3 & 18.4 & 0.5 & 11 & 0.17 & 45 & 2290 & 1 & 14 \\
HD 82943 c & 6.3 & 15.1 & 1.6 & 11 & 0.43 & 3.3 & 220 & 0.02 & 27 \\
HD 9826 c & 5.9 & 14.8 & 1.9 & 8.6 & 0.27 & 4.3 & 241 & 0.02 & 14 \\
HD 37124 c & 5.7 & 12.6 & 0.6 & 15 & 0.10 & 42 & 887 & 0.3 & 32 \\
HD 4203 b & 5.6 & 15.1 & 1.8 & 12 & 0.51 & 2.9 & 437 & 0.03 & 83 \\
HD 169830 b & 5.4 & 18.2 & 3.0 & 15 & 0.31 & 5.7 & 226 & 0.03 & 37 \\
HD 13931 b & 5.3 & 17.3 & 1.9 & 7.1 & 0.03 & 70 & 4444 & 2 & 47 \\
HD 4208 b & 5.3 & 19.7 & 0.8 & 15 & 0.06 & 64 & 830 & 0.2 & 34 \\
HD 188015 b & 4.9 & 17.7 & 1.4 & 11 & 0.17 & 15 & 462 & 0.08 & 51 \\
HD 141399 c & 4.1 & 19.4 & 1.3 & 15 & 0.05 & 12 & 202 & 0.04 & 37 \\
HD 134987 c & 4.1 & 8.2 & 0.9 & 11 & 0.15 & 33 & 5854 & 3 & 26 \\
HD 141399 e & 4.0 & 19.7 & 0.8 & 14 & 0.08 & 92 & 3445 & 5 & 37 \\
HD 92788 c & 4.0 & 20.0 & 2.8 & 13 & 0.31 & 19 & 8195 & 4 & 35 \\
HD 177830 b & 4.0 & 15.0 & 1.3 & 14 & 0.03 & 62 & 407 & 0.06 & 63 \\
HD 136925 b & 2.4 & 17.3 & 1.0 & 19 & 0.12 & 70 & 4597 & 5 & 47 \\
HD 120066 b & 1.5 & 3.7 & 3.1 & 4.3 & 0.83 & 6.1 & 23521 & 60 & 31 \\
GJ 876 e & 1.5 & 3.4 & 0.04 & 6.3 & 0.10 & 6.5 & 124 & 0.7 & 4.7 \\
\enddata
\end{deluxetable*}

\section{Acceleration of Joint Fitting with Linearization} \label{sec:linearization}

Although the methodology described in this work is an effective method to modeling the DR4 data for CLS (and will constitute our production pipeline), their are limitations to our injection experiment driven by computational complexity. In the sample presented here, there are several examples whose vectors $\theta^\text{sys}$ exceeded 50 parameters, leading to large convergence overheads. To make the problem more tractable, we employed a linearization scheme to reduce the dimensionality encountered by the sampler.

For a complete description of the method, we refer the reader to \citet{Wright2009}, where it is derived in depth, and summarize the critical ideas here. In short, a number of the parameters in $\theta^\text{sys}$ contribute to the observed RV and astrometric observables linearly, or can be transformed into a basis in which that contribution is linear, while the remaining parameters cannot (thus, ``non-linear''). Non-linear parameters must always be sampled in the usual fashion. However, for any given choice of those non-linear parameters, the linear parameters which maximize the likelihood can be solved for analytically via least-squares in one fell swoop.

While the linear nature of some parameters (e.g., $\gamma$) is obvious, a key component of this scheme is to linearize additional parameters via the coordinates
\begin{equation}
H = a \cos\omega \sin i_\text{orb},
\end{equation}
\begin{equation}
C = a \sin\omega \sin i_\text{orb},
\end{equation}
from which both the astrometric and RV signatures of the orbital motion can be expressed linearly, given sampled values of $i_\text{orb}$ and $\Omega$ (which fix the orbit's geometric projection onto the sky). The vector of transformed linear parameters $\vec\beta$ can be written as\footnote{We generalize \citet{Wright2009}'s single systemic velocity $v_0$ to a per-instrument offset $\gamma_k$, and additionally include a linear and quadratic RV drift term ($\dot\gamma$, $\ddot\gamma$), neither of which appear in the original formalism but are trivial to implement.}
\begin{equation}
\vec\beta = \left[H_1, C_1, \ldots, H_n, C_n, \gamma_1, \ldots, \gamma_{N_k}, \dot\gamma, \ddot\gamma, \alpha, \delta, \mu_\delta, \mu_\alpha, \varpi\right].
\end{equation}
Given the non-linear parameters, one constructs a design matrix $\boldsymbol{F}$ such that $\boldsymbol{F}\vec\beta$ reproduces the vector of both RV and astrometric observables. $\vec\beta$ can then be solved for directly and converted back to the standard basis, at a negligible fraction of the computational cost of sampling the linear parameters explicitly. Incorporating this into posterior sampling is then straightforward, with the log-likelihood of any proposed non-linear parameter vector evaluated along with the corresponding optimized $\vec\beta$. 

\begin{figure}
    \centering
    \includegraphics[width=0.5\linewidth]{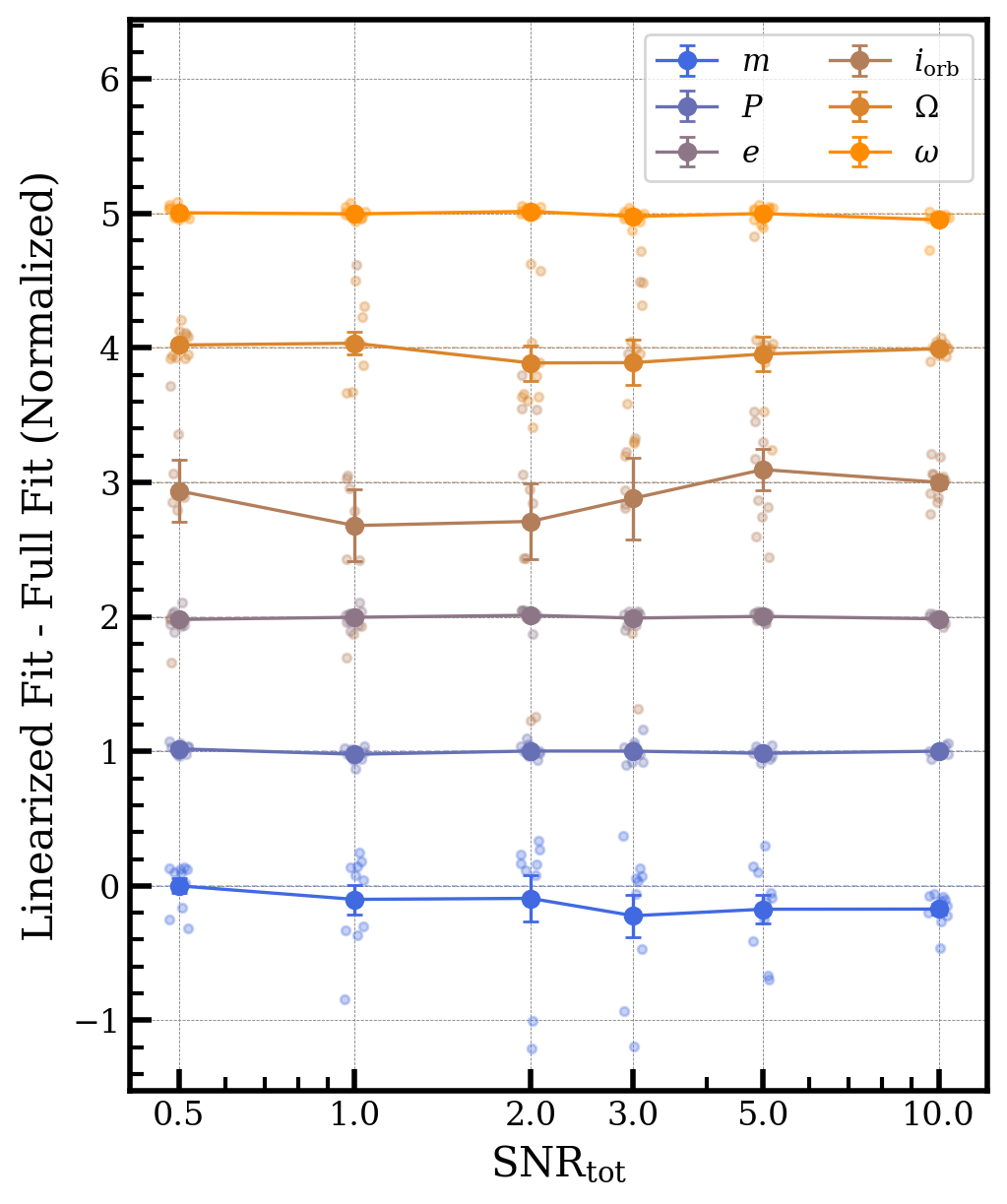}
    \caption{Difference between the posterior medians of a joint RV/astrometric fit with linearization active and inactive. Each parameter is given an arbitrary offset, and the differences between the two methods are normalized by the posterior widths (z-score). Any difference between the two bases is generally insignificant ($\ll 1\sigma$) for all the orbital parameters of interest.}
    \label{fig:linearization_test}
\end{figure}

We found that linearization does not significantly bias any of the companion orbital parameters. To demonstrate this, we ran a suite of simulations on a synthetic system. Around a solar-mass star, a companion was simulated with $m=2 M_\text{Jup}$, $e=0.2$, and the $P=500$ days, giving a strong astrometric signal well-resolved over the DR4 baseline. We tested a spectrum of $\snrast$ values between 0.5 and 10, wherein each run, we varied the system parallax (holding everything else constant) to get the desired $\snrast$. We generated a synthetic RV dataset for the system at typical observing cadence (we arbitrarily chose the epochs of the RV observations of HD~10697). At each $\snrast$, we generated ten orbits with varied the astrometric parameters. We fit the system with and without linearization active, and recorded the difference of the joint fit to each parameter (thereafter normalized by the run's uncertainty). We highlight the subset $( m, P, e, i_\text{orb},\Omega,\omega )$ in Figure \ref{fig:linearization_test}. The two fitting techniques agree almost perfectly, and any difference is well below the intrinsic measurement error, validating the use of partial linearization in this work.

Although the majority of the system parameters remain non-linear (most of the Keplerian basis, along with jitter terms), the improvement in dimensionality is substantial. For example: without linearization, a joint fit of the famed HD~219134 system (8 Keplerian signals, of which five we modeled jointly) requires an intensive 72 unique parameters. With linearization, the number of sampled parameters reduced to only 40. Lower dimensionality reduces the number of walkers necessary, as well as improving the absolute rate of convergence (i.e., the number of steps needed).

\section{Angular Momentum Deficit Equipartition} \label{sec:equipartition}

In this Section, we discuss a plausible width of the mutual inclination distribution using AMD equipartition. We make use of the canonical Poincar\'e coordinate system (e.g., \citealt{MurrayDermott}), and emphasize the actions
\begin{equation}
    \Lambda'=m\sqrt{GM_\star a},\qquad \Gamma'=\Lambda'(1-\sqrt{1-e^2}), \qquad Z'=\Lambda'\sqrt{1-e^2}(1-\cos i'),
\end{equation}
where $\Lambda'$ is the angular momentum associated with a circular and aligned orbit (the highest possible angular momentum state at fixed $a$). The action $\Gamma'$ corresponds to the AMD due to elongation into an eccentric orbit, and the action $Z'$ corresponds to the AMD acquired by tilting the orbit out of the reference plane. We note the critical distinction of $i'\neq i_\text{orb}$ from previous Sections --- the former is generally measured with reference to to the invariable plane of the planetary system, while the latter is unique to our line-of-sight view from Earth (and not generally of interest in the dynamical sense).

Assuming relatively small deviations from coplanar and circular orbits, to second order in $e'$ and $i'$, the AMD contributions conveniently decouple into
\begin{equation}
    \Gamma'\approx\Lambda'\frac{e^2}{2},\qquad
    Z'\approx\Lambda'\frac{{i'}^2}{2}.
\end{equation}
If 1:1 equipartition is satisfied, one expects $\langle\Gamma'\rangle\approx\langle Z'\rangle$, suggesting $\langle e^2\rangle\approx\langle {i'}^2\rangle$, an equivalence between the eccentricity and inclination (when expressed in radians). Returning to the mutual inclination (Equation \ref{eq:mutualinclination}), in the low-excitation limit,
\begin{equation}
    \langle\Phi^2\rangle
    \approx
    \langle {i'}_1^2\rangle+\langle {i'}_2^2\rangle - 2\langle i'_1 i'_2 \cos{(\Omega_1'-\Omega_2')}\rangle
    \approx
    2\langle {i'}^2\rangle,
\end{equation}
where we have taken $\langle\cos(\Omega_1'-\Omega_2')\rangle=0$ and assumed $\Omega_1'-\Omega_2'$ independent from inclination. This simple equipartition argument suggests
\begin{equation} \label{eq:equipartition}
    \sqrt{\langle\Phi^2\rangle}\approx\sqrt{2\langle e^2\rangle},
\end{equation}
a near equivalence of the root-mean-square values.

In N-body simulations of planetesimal swarms embedded in gaseous disks, equipartition manifests as a steady-state relation between the mean eccentricity and inclination, $\langle e\rangle\approx2\langle i'\rangle$, reflecting the preferential excitation of eccentricity \citep{Ida1992,Wetherill1980}. Accounting for the distinction between individual inclinations and pairwise mutual inclinations, this relation differs by approximately a factor of two from the 1:1 scaling adopted here. We note that the most energetic particles in those simulations can deviate substantially from this relation \citep{Shiidsuka1999}, and the characteristic eccentricities probed, $\langle e\rangle\sim10^{-2}$, are well below those observed for giant planets. Thus, although planetesimal-stirring calculations motivate a connection between eccentricity and inclination, they do not uniquely predict the mutual-inclination distribution of giant-planet systems. The strong planet--planet scattering invoked to explain the eccentricities of giant planets may instead produce a qualitatively different relationship between eccentricity and inclination \citep{Chatterjee2008,Juric2008}. We therefore adopt a 1:1 scaling as a demonstrative choice, without assuming it represents the equilibrium relation for giant-planet systems, although the relatively low mutual inclinations inferred for Kepler multiplanet systems are broadly consistent with the planetesimal-stirring picture \citep{Xie2016,Fabrycky2014}, as are constraints on debris-disk geometries \citep{Zawadzki2026,Daley2019}.

To determine the scale of $\alpha_\Phi$ and $\beta_\Phi$, we fit a Gamma distribution to the eccentricities of all the multi-giant systems in the CLS sample (including, but not limited to, the mutual inclination sample). This yielded a shape parameter for the eccentricity distribution of $\alpha_e\approx1.3$, and scale parameter $\beta_e\approx0.2$, which is consistent with the greater ensemble (specifically, with other multi-planet systems at large separations; \citealt{Stevenson2025}). Assuming the shape parameters are the same, we matched $\langle \Phi\rangle\approx\sqrt{2}\langle e \rangle\approx0.37$, granting $\beta_\Phi\approx16^\circ$.

\bibliography{main}{}
\bibliographystyle{aasjournalv7}

\end{document}